\documentclass[twocolumn,traditabstract,longauth]{aa}
\usepackage{amssymb}
\usepackage{mathptmx}
\usepackage{natbib}
\bibpunct{(}{)}{;}{a}{}{,}  
\usepackage[]{graphicx}
\usepackage{lscape}
\usepackage{txfonts}
\usepackage{url}

\def\setsymbol#1#2{\expandafter\def\csname #1\endcsname{#2}}
\def\getsymbol#1{\csname #1\endcsname}

\def\Planck{\textit{Planck}}

\def\all2013resultspapers{\nocite{planck2013-p01, planck2013-p02, planck2013-p02a, planck2013-p02d, planck2013-p02b, planck2013-p03, planck2013-p03c, planck2013-p03f, planck2013-p03d, planck2013-p03e, planck2013-p01a, planck2013-p06, planck2013-p03a, planck2013-pip88, planck2013-p08, planck2013-p11, planck2013-p12, planck2013-p13, planck2013-p14, planck2013-p15, planck2013-p05b, planck2013-p17, planck2013-p09, planck2013-p09a, planck2013-p20, planck2013-p19, planck2013-pipaberration, planck2013-p05, planck2013-p05a, planck2013-pip56, planck2013-p06b}}

\newbox\tablebox    \newdimen\tablewidth
\def\leaderfil{\leaders\hbox to 5pt{\hss.\hss}\hfil}
\def\endPlancktablewide{\tablewidth=\textwidth 
    $$\hss\copy\tablebox\hss$$
    \vskip-\lastskip\vskip -2pt}
\def\tablenote#1 #2\par{\begingroup \parindent=0.8em
    \abovedisplayshortskip=0pt\belowdisplayshortskip=0pt
    \noindent
    $$\hss\vbox{\hsize\tablewidth \hangindent=\parindent \hangafter=1 \noindent
    \hbox to \parindent{$^#1$\hss}\strut#2\strut\par}\hss$$
    \endgroup}
\def\doubleline{\vskip 3pt\hrule \vskip 1.5pt \hrule \vskip 5pt}

\def\L2{\ifmmode L_2\else $L_2$\fi}

\def\DeltaT{\ifmmode \Delta T\else $\Delta T$\fi}
\def\deltat{\ifmmode \Delta t\else $\Delta t$\fi}
\def\fknee{\ifmmode f_{\rm knee}\else $f_{\rm knee}$\fi}
\def\Fmax{\ifmmode F_{\rm max}\else $F_{\rm max}$\fi}
\def\solar{\ifmmode{\rm M}_{\mathord\odot}\else${\rm M}_{\mathord\odot}$\fi}
\def\Msolar{\ifmmode{\rm M}_{\mathord\odot}\else${\rm M}_{\mathord\odot}$\fi}
\def\Lsolar{\ifmmode{\rm L}_{\mathord\odot}\else${\rm L}_{\mathord\odot}$\fi}

\def\inv{\ifmmode^{-1}\else$^{-1}$\fi}
\def\mo{\ifmmode^{-1}\else$^{-1}$\fi}
\def\sup#1{\ifmmode ^{\rm #1}\else $^{\rm #1}$\fi}
\def\expo#1{\ifmmode \times 10^{#1}\else $\times 10^{#1}$\fi}
\def\,{\thinspace}
\def\lsim{\mathrel{\raise .4ex\hbox{\rlap{$<$}\lower 1.2ex\hbox{$\sim$}}}}
\def\gsim{\mathrel{\raise .4ex\hbox{\rlap{$>$}\lower 1.2ex\hbox{$\sim$}}}}

\def\simprop{\mathrel{\raise .4ex\hbox{\rlap{$\propto$}\lower 1.2ex\hbox{$\sim$}}}}
\def\deg{\ifmmode^\circ\else$^\circ$\fi}
\def\pdeg{\ifmmode $\setbox0=\hbox{$^{\circ}$}\rlap{\hskip.11\wd0 .}$^{\circ}
          \else \setbox0=\hbox{$^{\circ}$}\rlap{\hskip.11\wd0 .}$^{\circ}$\fi}
\def\arcs{\ifmmode {^{\scriptstyle\prime\prime}}
          \else $^{\scriptstyle\prime\prime}$\fi}
\def\arcm{\ifmmode {^{\scriptstyle\prime}}
          \else $^{\scriptstyle\prime}$\fi}
\newdimen\sa  \newdimen\sb
\def\parcs{\sa=.07em \sb=.03em
     \ifmmode \hbox{\rlap{.}}^{\scriptstyle\prime\kern -\sb\prime}\hbox{\kern -\sa}
     \else \rlap{.}$^{\scriptstyle\prime\kern -\sb\prime}$\kern -\sa\fi}
\def\parcm{\sa=.08em \sb=.03em
     \ifmmode \hbox{\rlap{.}\kern\sa}^{\scriptstyle\prime}\hbox{\kern-\sb}
     \else \rlap{.}\kern\sa$^{\scriptstyle\prime}$\kern-\sb\fi}
\def\ra[#1 #2 #3.#4]{#1\sup{h}#2\sup{m}#3\sup{s}\llap.#4}
\def\dec[#1 #2 #3.#4]{#1\deg#2\arcm#3\arcs\llap.#4}
\def\deco[#1 #2 #3]{#1\deg#2\arcm#3\arcs}
\def\rra[#1 #2]{#1\sup{h}#2\sup{m}}

\def\dots{\relax\ifmmode \ldots\else $\ldots$\fi}
\def\WHzsr{\ifmmode $W\,Hz\mo\,sr\mo$\else W\,Hz\mo\,sr\mo\fi}
\def\mHz{\ifmmode $\,mHz$\else \,mHz\fi}
\def\GHz{\ifmmode $\,GHz$\else \,GHz\fi}
\def\mKs{\ifmmode $\,mK\,s$^{1/2}\else \,mK\,s$^{1/2}$\fi}
\def\muKs{\ifmmode \,\mu$K\,s$^{1/2}\else \,$\mu$K\,s$^{1/2}$\fi}
\def\muKRJs{\ifmmode \,\mu$K$_{\rm RJ}$\,s$^{1/2}\else \,$\mu$K$_{\rm RJ}$\,s$^{1/2}$\fi}
\def\muKHz{\ifmmode \,\mu$K\,Hz$^{-1/2}\else \,$\mu$K\,Hz$^{-1/2}$\fi}
\def\MJysr{\ifmmode \,$MJy\,sr\mo$\else \,MJy\,sr\mo\fi}
\def\MJysrmK{\ifmmode \,$MJy\,sr\mo$\,mK$_{\rm CMB}\mo\else \,MJy\,sr\mo\,mK$_{\rm CMB}\mo$\fi}
\def\microns{\ifmmode \,\mu$m$\else \,$\mu$m\fi}

\def\muK{\ifmmode \,\mu$K$\else \,$\mu$\hbox{K}\fi}
\def\microK{\ifmmode \,\mu$K$\else \,$\mu$\hbox{K}\fi}
\def\muW{\ifmmode \,\mu$W$\else \,$\mu$\hbox{W}\fi}
\def\kms{\ifmmode $\,km\,s$^{-1}\else \,km\,s$^{-1}$\fi}
\def\kmsMpc{\ifmmode $\,\kms\,Mpc\mo$\else \,\kms\,Mpc\mo\fi}
\setsymbol{LFI:center:frequency:70GHz:units}{70.3\,GHz}
\setsymbol{LFI:center:frequency:44GHz:units}{44.1\,GHz}
\setsymbol{LFI:center:frequency:30GHz:units}{28.5\,GHz}

\setsymbol{LFI:center:frequency:70GHz}{70.3}
\setsymbol{LFI:center:frequency:44GHz}{44.1}
\setsymbol{LFI:center:frequency:30GHz}{28.5}

\setsymbol{LFI:center:frequency:LFI18:Rad:M:units}{71.7\GHz}
\setsymbol{LFI:center:frequency:LFI19:Rad:M:units}{67.5\GHz}
\setsymbol{LFI:center:frequency:LFI20:Rad:M:units}{69.2\GHz}
\setsymbol{LFI:center:frequency:LFI21:Rad:M:units}{70.4\GHz}
\setsymbol{LFI:center:frequency:LFI22:Rad:M:units}{71.5\GHz}
\setsymbol{LFI:center:frequency:LFI23:Rad:M:units}{70.8\GHz}
\setsymbol{LFI:center:frequency:LFI24:Rad:M:units}{44.4\GHz}
\setsymbol{LFI:center:frequency:LFI25:Rad:M:units}{44.0\GHz}
\setsymbol{LFI:center:frequency:LFI26:Rad:M:units}{43.9\GHz}
\setsymbol{LFI:center:frequency:LFI27:Rad:M:units}{28.3\GHz}
\setsymbol{LFI:center:frequency:LFI28:Rad:M:units}{28.8\GHz}
\setsymbol{LFI:center:frequency:LFI18:Rad:S:units}{70.1\GHz}
\setsymbol{LFI:center:frequency:LFI19:Rad:S:units}{69.6\GHz}
\setsymbol{LFI:center:frequency:LFI20:Rad:S:units}{69.5\GHz}
\setsymbol{LFI:center:frequency:LFI21:Rad:S:units}{69.5\GHz}
\setsymbol{LFI:center:frequency:LFI22:Rad:S:units}{72.8\GHz}
\setsymbol{LFI:center:frequency:LFI23:Rad:S:units}{71.3\GHz}
\setsymbol{LFI:center:frequency:LFI24:Rad:S:units}{44.1\GHz}
\setsymbol{LFI:center:frequency:LFI25:Rad:S:units}{44.1\GHz}
\setsymbol{LFI:center:frequency:LFI26:Rad:S:units}{44.1\GHz}
\setsymbol{LFI:center:frequency:LFI27:Rad:S:units}{28.5\GHz}
\setsymbol{LFI:center:frequency:LFI28:Rad:S:units}{28.2\GHz}

\setsymbol{LFI:center:frequency:LFI18:Rad:M}{71.7}
\setsymbol{LFI:center:frequency:LFI19:Rad:M}{67.5}
\setsymbol{LFI:center:frequency:LFI20:Rad:M}{69.2}
\setsymbol{LFI:center:frequency:LFI21:Rad:M}{70.4}
\setsymbol{LFI:center:frequency:LFI22:Rad:M}{71.5}
\setsymbol{LFI:center:frequency:LFI23:Rad:M}{70.8}
\setsymbol{LFI:center:frequency:LFI24:Rad:M}{44.4}
\setsymbol{LFI:center:frequency:LFI25:Rad:M}{44.0}
\setsymbol{LFI:center:frequency:LFI26:Rad:M}{43.9}
\setsymbol{LFI:center:frequency:LFI27:Rad:M}{28.3}
\setsymbol{LFI:center:frequency:LFI28:Rad:M}{28.8}
\setsymbol{LFI:center:frequency:LFI18:Rad:S}{70.1}
\setsymbol{LFI:center:frequency:LFI19:Rad:S}{69.6}
\setsymbol{LFI:center:frequency:LFI20:Rad:S}{69.5}
\setsymbol{LFI:center:frequency:LFI21:Rad:S}{69.5}
\setsymbol{LFI:center:frequency:LFI22:Rad:S}{72.8}
\setsymbol{LFI:center:frequency:LFI23:Rad:S}{71.3}
\setsymbol{LFI:center:frequency:LFI24:Rad:S}{44.1}
\setsymbol{LFI:center:frequency:LFI25:Rad:S}{44.1}
\setsymbol{LFI:center:frequency:LFI26:Rad:S}{44.1}
\setsymbol{LFI:center:frequency:LFI27:Rad:S}{28.5}
\setsymbol{LFI:center:frequency:LFI28:Rad:S}{28.2}

\setsymbol{LFI:white:noise:sensitivity:70GHz:units}{134.7\muKs}
\setsymbol{LFI:white:noise:sensitivity:44GHz:units}{164.7\muKs}
\setsymbol{LFI:white:noise:sensitivity:30GHz:units}{143.4\muKs}

\setsymbol{LFI:white:noise:sensitivity:70GHz}{134.7}
\setsymbol{LFI:white:noise:sensitivity:44GHz}{164.7}
\setsymbol{LFI:white:noise:sensitivity:30GHz}{143.4}

\setsymbol{LFI:white:noise:sensitivity:LFI18:Rad:M:units}{512.0\muKs}
\setsymbol{LFI:white:noise:sensitivity:LFI19:Rad:M:units}{581.4\muKs}
\setsymbol{LFI:white:noise:sensitivity:LFI20:Rad:M:units}{590.8\muKs}
\setsymbol{LFI:white:noise:sensitivity:LFI21:Rad:M:units}{455.2\muKs}
\setsymbol{LFI:white:noise:sensitivity:LFI22:Rad:M:units}{492.0\muKs}
\setsymbol{LFI:white:noise:sensitivity:LFI23:Rad:M:units}{507.7\muKs}
\setsymbol{LFI:white:noise:sensitivity:LFI24:Rad:M:units}{462.2\muKs}
\setsymbol{LFI:white:noise:sensitivity:LFI25:Rad:M:units}{413.6\muKs}
\setsymbol{LFI:white:noise:sensitivity:LFI26:Rad:M:units}{478.6\muKs}
\setsymbol{LFI:white:noise:sensitivity:LFI27:Rad:M:units}{277.7\muKs}
\setsymbol{LFI:white:noise:sensitivity:LFI28:Rad:M:units}{312.3\muKs}
\setsymbol{LFI:white:noise:sensitivity:LFI18:Rad:S:units}{465.7\muKs}
\setsymbol{LFI:white:noise:sensitivity:LFI19:Rad:S:units}{555.6\muKs}
\setsymbol{LFI:white:noise:sensitivity:LFI20:Rad:S:units}{623.2\muKs}
\setsymbol{LFI:white:noise:sensitivity:LFI21:Rad:S:units}{564.1\muKs}
\setsymbol{LFI:white:noise:sensitivity:LFI22:Rad:S:units}{534.4\muKs}
\setsymbol{LFI:white:noise:sensitivity:LFI23:Rad:S:units}{542.4\muKs}
\setsymbol{LFI:white:noise:sensitivity:LFI24:Rad:S:units}{399.2\muKs}
\setsymbol{LFI:white:noise:sensitivity:LFI25:Rad:S:units}{392.6\muKs}
\setsymbol{LFI:white:noise:sensitivity:LFI26:Rad:S:units}{418.6\muKs}
\setsymbol{LFI:white:noise:sensitivity:LFI27:Rad:S:units}{302.9\muKs}
\setsymbol{LFI:white:noise:sensitivity:LFI28:Rad:S:units}{285.3\muKs}

\setsymbol{LFI:white:noise:sensitivity:LFI18:Rad:M}{512.0}
\setsymbol{LFI:white:noise:sensitivity:LFI19:Rad:M}{581.4}
\setsymbol{LFI:white:noise:sensitivity:LFI20:Rad:M}{590.8}
\setsymbol{LFI:white:noise:sensitivity:LFI21:Rad:M}{455.2}
\setsymbol{LFI:white:noise:sensitivity:LFI22:Rad:M}{492.0}
\setsymbol{LFI:white:noise:sensitivity:LFI23:Rad:M}{507.7}
\setsymbol{LFI:white:noise:sensitivity:LFI24:Rad:M}{462.2}
\setsymbol{LFI:white:noise:sensitivity:LFI25:Rad:M}{413.6}
\setsymbol{LFI:white:noise:sensitivity:LFI26:Rad:M}{478.6}
\setsymbol{LFI:white:noise:sensitivity:LFI27:Rad:M}{277.7}
\setsymbol{LFI:white:noise:sensitivity:LFI28:Rad:M}{312.3}
\setsymbol{LFI:white:noise:sensitivity:LFI18:Rad:S}{465.7}
\setsymbol{LFI:white:noise:sensitivity:LFI19:Rad:S}{555.6}
\setsymbol{LFI:white:noise:sensitivity:LFI20:Rad:S}{623.2}
\setsymbol{LFI:white:noise:sensitivity:LFI21:Rad:S}{564.1}
\setsymbol{LFI:white:noise:sensitivity:LFI22:Rad:S}{534.4}
\setsymbol{LFI:white:noise:sensitivity:LFI23:Rad:S}{542.4}
\setsymbol{LFI:white:noise:sensitivity:LFI24:Rad:S}{399.2}
\setsymbol{LFI:white:noise:sensitivity:LFI25:Rad:S}{392.6}
\setsymbol{LFI:white:noise:sensitivity:LFI26:Rad:S}{418.6}
\setsymbol{LFI:white:noise:sensitivity:LFI27:Rad:S}{302.9}
\setsymbol{LFI:white:noise:sensitivity:LFI28:Rad:S}{285.3}

\setsymbol{LFI:knee:frequency:70GHz:units}{29.5\mHz}
\setsymbol{LFI:knee:frequency:44GHz:units}{56.2\mHz}
\setsymbol{LFI:knee:frequency:30GHz:units}{113.7\mHz}

\setsymbol{LFI:knee:frequency:70GHz}{29.5}
\setsymbol{LFI:knee:frequency:44GHz}{56.2}
\setsymbol{LFI:knee:frequency:30GHz}{113.7}

\setsymbol{LFI:knee:frequency:LFI18:Rad:M:units}{16.3\mHz}
\setsymbol{LFI:knee:frequency:LFI19:Rad:M:units}{15.1\mHz}
\setsymbol{LFI:knee:frequency:LFI20:Rad:M:units}{18.7\mHz}
\setsymbol{LFI:knee:frequency:LFI21:Rad:M:units}{37.2\mHz}
\setsymbol{LFI:knee:frequency:LFI22:Rad:M:units}{12.7\mHz}
\setsymbol{LFI:knee:frequency:LFI23:Rad:M:units}{34.6\mHz}
\setsymbol{LFI:knee:frequency:LFI24:Rad:M:units}{46.2\mHz}
\setsymbol{LFI:knee:frequency:LFI25:Rad:M:units}{24.9\mHz}
\setsymbol{LFI:knee:frequency:LFI26:Rad:M:units}{67.6\mHz}
\setsymbol{LFI:knee:frequency:LFI27:Rad:M:units}{187.4\mHz}
\setsymbol{LFI:knee:frequency:LFI28:Rad:M:units}{122.2\mHz}
\setsymbol{LFI:knee:frequency:LFI18:Rad:S:units}{17.7\mHz}
\setsymbol{LFI:knee:frequency:LFI19:Rad:S:units}{22.0\mHz}
\setsymbol{LFI:knee:frequency:LFI20:Rad:S:units}{8.7\mHz}
\setsymbol{LFI:knee:frequency:LFI21:Rad:S:units}{25.9\mHz}
\setsymbol{LFI:knee:frequency:LFI22:Rad:S:units}{15.8\mHz}
\setsymbol{LFI:knee:frequency:LFI23:Rad:S:units}{129.8\mHz}
\setsymbol{LFI:knee:frequency:LFI24:Rad:S:units}{100.9\mHz}
\setsymbol{LFI:knee:frequency:LFI25:Rad:S:units}{38.9\mHz}
\setsymbol{LFI:knee:frequency:LFI26:Rad:S:units}{58.9\mHz}
\setsymbol{LFI:knee:frequency:LFI27:Rad:S:units}{104.4\mHz}
\setsymbol{LFI:knee:frequency:LFI28:Rad:S:units}{40.7\mHz}

\setsymbol{LFI:knee:frequency:LFI18:Rad:M}{16.3}
\setsymbol{LFI:knee:frequency:LFI19:Rad:M}{15.1}
\setsymbol{LFI:knee:frequency:LFI20:Rad:M}{18.7}
\setsymbol{LFI:knee:frequency:LFI21:Rad:M}{37.2}
\setsymbol{LFI:knee:frequency:LFI22:Rad:M}{12.7}
\setsymbol{LFI:knee:frequency:LFI23:Rad:M}{34.6}
\setsymbol{LFI:knee:frequency:LFI24:Rad:M}{46.2}
\setsymbol{LFI:knee:frequency:LFI25:Rad:M}{24.9}
\setsymbol{LFI:knee:frequency:LFI26:Rad:M}{67.6}
\setsymbol{LFI:knee:frequency:LFI27:Rad:M}{187.4}
\setsymbol{LFI:knee:frequency:LFI28:Rad:M}{122.2}
\setsymbol{LFI:knee:frequency:LFI18:Rad:S}{17.7}
\setsymbol{LFI:knee:frequency:LFI19:Rad:S}{22.0}
\setsymbol{LFI:knee:frequency:LFI20:Rad:S}{8.7}
\setsymbol{LFI:knee:frequency:LFI21:Rad:S}{25.9}
\setsymbol{LFI:knee:frequency:LFI22:Rad:S}{15.8}
\setsymbol{LFI:knee:frequency:LFI23:Rad:S}{129.8}
\setsymbol{LFI:knee:frequency:LFI24:Rad:S}{100.9}
\setsymbol{LFI:knee:frequency:LFI25:Rad:S}{38.9}
\setsymbol{LFI:knee:frequency:LFI26:Rad:S}{58.9}
\setsymbol{LFI:knee:frequency:LFI27:Rad:S}{104.4}
\setsymbol{LFI:knee:frequency:LFI28:Rad:S}{40.7}

\setsymbol{LFI:slope:70GHz:units}{$-1.03$\mHz}
\setsymbol{LFI:slope:44GHz:units}{$-0.89$\mHz}
\setsymbol{LFI:slope:30GHz:units}{$-0.87$\mHz}

\setsymbol{LFI:slope:70GHz}{$-1.03$}
\setsymbol{LFI:slope:44GHz}{$-0.89$}
\setsymbol{LFI:slope:30GHz}{$-0.87$}

\setsymbol{LFI:slope:LFI18:Rad:M:units}{$-1.04$\mHz}
\setsymbol{LFI:slope:LFI19:Rad:M:units}{$-1.09$\mHz}
\setsymbol{LFI:slope:LFI20:Rad:M:units}{$-0.69$\mHz}
\setsymbol{LFI:slope:LFI21:Rad:M:units}{$-1.56$\mHz}
\setsymbol{LFI:slope:LFI22:Rad:M:units}{$-1.01$\mHz}
\setsymbol{LFI:slope:LFI23:Rad:M:units}{$-0.96$\mHz}
\setsymbol{LFI:slope:LFI24:Rad:M:units}{$-0.83$\mHz}
\setsymbol{LFI:slope:LFI25:Rad:M:units}{$-0.91$\mHz}
\setsymbol{LFI:slope:LFI26:Rad:M:units}{$-0.95$\mHz}
\setsymbol{LFI:slope:LFI27:Rad:M:units}{$-0.87$\mHz}
\setsymbol{LFI:slope:LFI28:Rad:M:units}{$-0.88$\mHz}
\setsymbol{LFI:slope:LFI18:Rad:S:units}{$-1.15$\mHz}
\setsymbol{LFI:slope:LFI19:Rad:S:units}{$-1.00$\mHz}
\setsymbol{LFI:slope:LFI20:Rad:S:units}{$-0.95$\mHz}
\setsymbol{LFI:slope:LFI21:Rad:S:units}{$-0.92$\mHz}
\setsymbol{LFI:slope:LFI22:Rad:S:units}{$-1.01$\mHz}
\setsymbol{LFI:slope:LFI23:Rad:S:units}{$-0.95$\mHz}
\setsymbol{LFI:slope:LFI24:Rad:S:units}{$-0.73$\mHz}
\setsymbol{LFI:slope:LFI25:Rad:S:units}{$-1.16$\mHz}
\setsymbol{LFI:slope:LFI26:Rad:S:units}{$-0.79$\mHz}
\setsymbol{LFI:slope:LFI27:Rad:S:units}{$-0.82$\mHz}
\setsymbol{LFI:slope:LFI28:Rad:S:units}{$-0.91$\mHz}

\setsymbol{LFI:slope:LFI18:Rad:M}{$-1.04$}
\setsymbol{LFI:slope:LFI19:Rad:M}{$-1.09$}
\setsymbol{LFI:slope:LFI20:Rad:M}{$-0.69$}
\setsymbol{LFI:slope:LFI21:Rad:M}{$-1.56$}
\setsymbol{LFI:slope:LFI22:Rad:M}{$-1.01$}
\setsymbol{LFI:slope:LFI23:Rad:M}{$-0.96$}
\setsymbol{LFI:slope:LFI24:Rad:M}{$-0.83$}
\setsymbol{LFI:slope:LFI25:Rad:M}{$-0.91$}
\setsymbol{LFI:slope:LFI26:Rad:M}{$-0.95$}
\setsymbol{LFI:slope:LFI27:Rad:M}{$-0.87$}
\setsymbol{LFI:slope:LFI28:Rad:M}{$-0.88$}
\setsymbol{LFI:slope:LFI18:Rad:S}{$-1.15$}
\setsymbol{LFI:slope:LFI19:Rad:S}{$-1.00$}
\setsymbol{LFI:slope:LFI20:Rad:S}{$-0.95$}
\setsymbol{LFI:slope:LFI21:Rad:S}{$-0.92$}
\setsymbol{LFI:slope:LFI22:Rad:S}{$-1.01$}
\setsymbol{LFI:slope:LFI23:Rad:S}{$-0.95$}
\setsymbol{LFI:slope:LFI24:Rad:S}{$-0.73$}
\setsymbol{LFI:slope:LFI25:Rad:S}{$-1.16$}
\setsymbol{LFI:slope:LFI26:Rad:S}{$-0.79$}
\setsymbol{LFI:slope:LFI27:Rad:S}{$-0.82$}
\setsymbol{LFI:slope:LFI28:Rad:S}{$-0.91$}

\setsymbol{LFI:FWHM:70GHz:units}{13\parcm01}
\setsymbol{LFI:FWHM:44GHz:units}{27\parcm92}
\setsymbol{LFI:FWHM:30GHz:units}{32\parcm65}

\setsymbol{LFI:FWHM:70GHz}{13.01}
\setsymbol{LFI:FWHM:44GHz}{27.92}
\setsymbol{LFI:FWHM:30GHz}{32.65}

\setsymbol{LFI:FWHM:LFI18:units}{13\parcm39}
\setsymbol{LFI:FWHM:LFI19:units}{13\parcm01}
\setsymbol{LFI:FWHM:LFI20:units}{12\parcm75}
\setsymbol{LFI:FWHM:LFI21:units}{12\parcm74}
\setsymbol{LFI:FWHM:LFI22:units}{12\parcm87}
\setsymbol{LFI:FWHM:LFI23:units}{13\parcm27}
\setsymbol{LFI:FWHM:LFI24:units}{22\parcm98}
\setsymbol{LFI:FWHM:LFI25:units}{30\parcm46}
\setsymbol{LFI:FWHM:LFI26:units}{30\parcm31}
\setsymbol{LFI:FWHM:LFI27:units}{32\parcm65}
\setsymbol{LFI:FWHM:LFI28:units}{32\parcm66}

\setsymbol{LFI:FWHM:LFI18}{13.39}
\setsymbol{LFI:FWHM:LFI19}{13.01}
\setsymbol{LFI:FWHM:LFI20}{12.75}
\setsymbol{LFI:FWHM:LFI21}{12.74}
\setsymbol{LFI:FWHM:LFI22}{12.87}
\setsymbol{LFI:FWHM:LFI23}{13.27}
\setsymbol{LFI:FWHM:LFI24}{22.98}
\setsymbol{LFI:FWHM:LFI25}{30.46}
\setsymbol{LFI:FWHM:LFI26}{30.31}
\setsymbol{LFI:FWHM:LFI27}{32.65}
\setsymbol{LFI:FWHM:LFI28}{32.66}

\setsymbol{LFI:FWHM:uncertainty:LFI18:units}{0.170\arcm}
\setsymbol{LFI:FWHM:uncertainty:LFI19:units}{0.174\arcm}
\setsymbol{LFI:FWHM:uncertainty:LFI20:units}{0.170\arcm}
\setsymbol{LFI:FWHM:uncertainty:LFI21:units}{0.156\arcm}
\setsymbol{LFI:FWHM:uncertainty:LFI22:units}{0.164\arcm}
\setsymbol{LFI:FWHM:uncertainty:LFI23:units}{0.171\arcm}
\setsymbol{LFI:FWHM:uncertainty:LFI24:units}{0.652\arcm}
\setsymbol{LFI:FWHM:uncertainty:LFI25:units}{1.075\arcm}
\setsymbol{LFI:FWHM:uncertainty:LFI26:units}{1.131\arcm}
\setsymbol{LFI:FWHM:uncertainty:LFI27:units}{1.266\arcm}
\setsymbol{LFI:FWHM:uncertainty:LFI28:units}{1.287\arcm}

\setsymbol{LFI:FWHM:uncertainty:LFI18}{0.170}
\setsymbol{LFI:FWHM:uncertainty:LFI19}{0.174}
\setsymbol{LFI:FWHM:uncertainty:LFI20}{0.170}
\setsymbol{LFI:FWHM:uncertainty:LFI21}{0.156}
\setsymbol{LFI:FWHM:uncertainty:LFI22}{0.164}
\setsymbol{LFI:FWHM:uncertainty:LFI23}{0.171}
\setsymbol{LFI:FWHM:uncertainty:LFI24}{0.652}
\setsymbol{LFI:FWHM:uncertainty:LFI25}{1.075}
\setsymbol{LFI:FWHM:uncertainty:LFI26}{1.131}
\setsymbol{LFI:FWHM:uncertainty:LFI27}{1.266}
\setsymbol{LFI:FWHM:uncertainty:LFI28}{1.287}

\setsymbol{HFI:center:frequency:100GHz:units}{100\,GHz}
\setsymbol{HFI:center:frequency:143GHz:units}{143\,GHz}
\setsymbol{HFI:center:frequency:217GHz:units}{217\,GHz}
\setsymbol{HFI:center:frequency:353GHz:units}{353\,GHz}
\setsymbol{HFI:center:frequency:545GHz:units}{545\,GHz}
\setsymbol{HFI:center:frequency:857GHz:units}{857\,GHz}

\setsymbol{HFI:center:frequency:100GHz}{100}
\setsymbol{HFI:center:frequency:143GHz}{143}
\setsymbol{HFI:center:frequency:217GHz}{217}
\setsymbol{HFI:center:frequency:353GHz}{353}
\setsymbol{HFI:center:frequency:545GHz}{545}
\setsymbol{HFI:center:frequency:857GHz}{857}

\setsymbol{HFI:Ndetectors:100GHz}{8}
\setsymbol{HFI:Ndetectors:143GHz}{11}
\setsymbol{HFI:Ndetectors:217GHz}{12}
\setsymbol{HFI:Ndetectors:353GHz}{12}
\setsymbol{HFI:Ndetectors:545GHz}{3}
\setsymbol{HFI:Ndetectors:857GHz}{4}

\setsymbol{HFI:FWHM:Maps:100GHz:units}{9\parcm88}
\setsymbol{HFI:FWHM:Maps:143GHz:units}{7\parcm18}
\setsymbol{HFI:FWHM:Maps:217GHz:units}{4\parcm87}
\setsymbol{HFI:FWHM:Maps:353GHz:units}{4\parcm65}
\setsymbol{HFI:FWHM:Maps:545GHz:units}{4\parcm72}
\setsymbol{HFI:FWHM:Maps:857GHz:units}{4\parcm39}
\setsymbol{HFI:FWHM:Maps:100GHz}{9.88}
\setsymbol{HFI:FWHM:Maps:143GHz}{7.18}
\setsymbol{HFI:FWHM:Maps:217GHz}{4.87}
\setsymbol{HFI:FWHM:Maps:353GHz}{4.65}
\setsymbol{HFI:FWHM:Maps:545GHz}{4.72}
\setsymbol{HFI:FWHM:Maps:857GHz}{4.39}

\setsymbol{HFI:beam:ellipticity:Maps:100GHz}{1.15}
\setsymbol{HFI:beam:ellipticity:Maps:143GHz}{1.01}
\setsymbol{HFI:beam:ellipticity:Maps:217GHz}{1.06}
\setsymbol{HFI:beam:ellipticity:Maps:353GHz}{1.05}
\setsymbol{HFI:beam:ellipticity:Maps:545GHz}{1.14}
\setsymbol{HFI:beam:ellipticity:Maps:857GHz}{1.19}

\setsymbol{HFI:FWHM:Mars:100GHz:units}{9\parcm37}
\setsymbol{HFI:FWHM:Mars:143GHz:units}{7\parcm04}
\setsymbol{HFI:FWHM:Mars:217GHz:units}{4\parcm68}
\setsymbol{HFI:FWHM:Mars:353GHz:units}{4\parcm43}
\setsymbol{HFI:FWHM:Mars:545GHz:units}{3\parcm80}
\setsymbol{HFI:FWHM:Mars:857GHz:units}{3\parcm67}

\setsymbol{HFI:FWHM:Mars:100GHz}{9.37}
\setsymbol{HFI:FWHM:Mars:143GHz}{7.04}
\setsymbol{HFI:FWHM:Mars:217GHz}{4.68}
\setsymbol{HFI:FWHM:Mars:353GHz}{4.43}
\setsymbol{HFI:FWHM:Mars:545GHz}{3.80}
\setsymbol{HFI:FWHM:Mars:857GHz}{3.67}

\setsymbol{HFI:beam:ellipticity:Mars:100GHz}{1.18}
\setsymbol{HFI:beam:ellipticity:Mars:143GHz}{1.03}
\setsymbol{HFI:beam:ellipticity:Mars:217GHz}{1.14}
\setsymbol{HFI:beam:ellipticity:Mars:353GHz}{1.09}
\setsymbol{HFI:beam:ellipticity:Mars:545GHz}{1.25}
\setsymbol{HFI:beam:ellipticity:Mars:857GHz}{1.03}

\setsymbol{HFI:CMB:relative:calibration:100GHz}{$\lsim 1\%$}
\setsymbol{HFI:CMB:relative:calibration:143GHz}{$\lsim 1\%$}
\setsymbol{HFI:CMB:relative:calibration:217GHz}{$\lsim 1\%$}
\setsymbol{HFI:CMB:relative:calibration:353GHz}{$\lsim 1\%$}
\setsymbol{HFI:CMB:relative:calibration:545GHz}{}
\setsymbol{HFI:CMB:relative:calibration:857GHz}{}

\setsymbol{HFI:CMB:absolute:calibration:100GHz}{$\lsim 2\%$}
\setsymbol{HFI:CMB:absolute:calibration:143GHz}{$\lsim 2\%$}
\setsymbol{HFI:CMB:absolute:calibration:217GHz}{$\lsim 2\%$}
\setsymbol{HFI:CMB:absolute:calibration:353GHz}{$\lsim 2\%$}
\setsymbol{HFI:CMB:absolute:calibration:545GHz}{}
\setsymbol{HFI:CMB:absolute:calibration:857GHz}{}

\setsymbol{HFI:FIRAS:gain:calibration:accuracy:statistical:100GHz}{}
\setsymbol{HFI:FIRAS:gain:calibration:accuracy:statistical:143GHz}{}
\setsymbol{HFI:FIRAS:gain:calibration:accuracy:statistical:217GHz}{}
\setsymbol{HFI:FIRAS:gain:calibration:accuracy:statistical:353GHz}{2.5\%}
\setsymbol{HFI:FIRAS:gain:calibration:accuracy:statistical:545GHz}{1\%}
\setsymbol{HFI:FIRAS:gain:calibration:accuracy:statistical:857GHz}{0.5\%}

\setsymbol{HFI:FIRAS:gain:calibration:accuracy:systematic:100GHz}{}
\setsymbol{HFI:FIRAS:gain:calibration:accuracy:systematic:143GHz}{}
\setsymbol{HFI:FIRAS:gain:calibration:accuracy:systematic:217GHz}{}
\setsymbol{HFI:FIRAS:gain:calibration:accuracy:systematic:353GHz}{}
\setsymbol{HFI:FIRAS:gain:calibration:accuracy:systematic:545GHz}{7\%}
\setsymbol{HFI:FIRAS:gain:calibration:accuracy:systematic:857GHz}{7\%}

\setsymbol{HFI:FIRAS:zero:point:accuracy:100GHz:units}{0.8\MJysr}
\setsymbol{HFI:FIRAS:zero:point:accuracy:143GHz:units}{}
\setsymbol{HFI:FIRAS:zero:point:accuracy:217GHz:units}{}
\setsymbol{HFI:FIRAS:zero:point:accuracy:353GHz:units}{1.4\MJysr}
\setsymbol{HFI:FIRAS:zero:point:accuracy:545GHz:units}{2.2\MJysr}
\setsymbol{HFI:FIRAS:zero:point:accuracy:857GHz:units}{1.7\MJysr}

\setsymbol{HFI:FIRAS:zero:point:accuracy:100GHz}{0.8}
\setsymbol{HFI:FIRAS:zero:point:accuracy:143GHz}{}
\setsymbol{HFI:FIRAS:zero:point:accuracy:217GHz}{}
\setsymbol{HFI:FIRAS:zero:point:accuracy:353GHz}{1.4}
\setsymbol{HFI:FIRAS:zero:point:accuracy:545GHz}{2.2}
\setsymbol{HFI:FIRAS:zero:point:accuracy:857GHz}{1.7}

\setsymbol{HFI:unit:conversion:100GHz:units}{0.2415\MJysrmK}
\setsymbol{HFI:unit:conversion:143GHz:units}{0.3694\MJysrmK}
\setsymbol{HFI:unit:conversion:217GHz:units}{0.4811\MJysrmK}
\setsymbol{HFI:unit:conversion:353GHz:units}{0.2883\MJysrmK}
\setsymbol{HFI:unit:conversion:545GHz:units}{0.05826\MJysrmK}
\setsymbol{HFI:unit:conversion:857GHz:units}{0.002238\MJysrmK}

\setsymbol{HFI:unit:conversion:100GHz}{0.2415}
\setsymbol{HFI:unit:conversion:143GHz}{0.3694}
\setsymbol{HFI:unit:conversion:217GHz}{0.4811}
\setsymbol{HFI:unit:conversion:353GHz}{0.2883}
\setsymbol{HFI:unit:conversion:545GHz}{0.05826}
\setsymbol{HFI:unit:conversion:857GHz}{0.002238}

\setsymbol{HFI:colour:correction:alpha=-2:V1.01:100GHz}{0.9893}
\setsymbol{HFI:colour:correction:alpha=-2:V1.01:143GHz}{0.9759}
\setsymbol{HFI:colour:correction:alpha=-2:V1.01:217GHz}{1.0007}
\setsymbol{HFI:colour:correction:alpha=-2:V1.01:353GHz}{1.0028}
\setsymbol{HFI:colour:correction:alpha=-2:V1.01:545GHz}{1.0019}
\setsymbol{HFI:colour:correction:alpha=-2:V1.01:857GHz}{0.9889}

\setsymbol{HFI:colour:correction:alpha=0:V1.01:100GHz}{1.0008}
\setsymbol{HFI:colour:correction:alpha=0:V1.01:143GHz}{1.0148}
\setsymbol{HFI:colour:correction:alpha=0:V1.01:217GHz}{0.9909}
\setsymbol{HFI:colour:correction:alpha=0:V1.01:353GHz}{0.9888}
\setsymbol{HFI:colour:correction:alpha=0:V1.01:545GHz}{0.9878}
\setsymbol{HFI:colour:correction:alpha=0:V1.01:857GHz}{1.0014}

\providecommand{\sorthelp}[1]{}

\begin{document}

\title{\Planck's Dusty GEMS: Gravitationally lensed high-redshift galaxies discovered with the \Planck\ survey
\thanks{Based on observations collected with the \textit{Herschel\/} and
 \Planck\ satellites, IRAM, SMA, JCMT, CFHT, and the VLT.}
}
\author{R.~Ca{\~n}ameras\inst{1,2}
\and
N.~P.~H.~Nesvadba\inst{2,1}~\thanks{Corresponding author:
N.~Nesvadba, nicole.nesvadba@ias.u-psud.fr}
\and
D.~Guery\inst{1,2}
\and
T.~McKenzie\inst{4}
\and
S.~K\"onig\inst{5}
\and
G.~Petitpas\inst{6}
\and
H.~Dole\inst{1,2,3}
\and
B.~Frye\inst{7}
\and
I.~Flores-Cacho\inst{8,9}
\and
L.~Montier\inst{8,9}
\and
M.~Negrello\inst{10}
\and
A.~Beelen\inst{1,2}
\and
F.~Boone\inst{9,10}
\and
D.~Dicken\inst{1,2,11}
\and
G.~Lagache\inst{1,2,12}
\and
E.~Le~Floc'h\inst{11}
\and
B.~Altieri\inst{13}
\and
M.~B{\'e}thermin\inst{14}
\and
R.~Chary\inst{15}
\and
G.~de~Zotti\inst{10}
\and
M.~Giard\inst{8,9}
\and
R.~Kneissl\inst{16,17}
\and
M.~Krips\inst{5}
\and
S.~Malhotra\inst{18}
\and
C.~Martinache\inst{1,2}
\and
A.~Omont\inst{19}
\and
E.~Pointecouteau\inst{8,9}
\and
J.-L.~Puget\inst{2,1}
\and
D.~Scott\inst{4}
\and
G.~Soucail\inst{8,9}
\and
I.~Valtchanov\inst{13}
\and
N.~Welikala\inst{20}
\and
L.~Yan\inst{13}}
\institute{Institut d'Astrophysique Spatiale, UMR8617, Universit\'e Paris-Sud, b\^at 121, Orsay, France
\and
CNRS, Orsay, France
\and 
Institut Universitaire de France
\and
Department of Physics \& Astronomy, University of British
66 Columbia, 6224 Agricultural Road, Vancouver, British Columbia, 58
Canada
\and
Institut de Radio Astronomie Millimétrique (IRAM), 300 rue de la Piscine, Domaine Universitaire, F-38406 Saint Martin d'Hères, France
\and 
Harvard-Smithsonian Center for Astrophysics, Cambridge, MA 02138, USA)
\and 
Steward Observatory, University of Arizona, Tucson, AZ 85721, USA
\and 
Universit\'e de Toulouse, UMS-OMP, IRAP, F-31028 Toulouse cedex 4, France
\and
CNRS, IRAP, 9 Av. colonel Roche, BP 44346, F-31028 Toulouse cedex 4, France
\and
INAF, Osservatorio Astronomico di Padova, Vicolo dell'Osservatorio 5, I-35122 Padova, Italy
\and 
CEA-Saclay, F-91191 Gif-sur-Yvette, France
\and
Aix Marseille Universit\'e, CNRS, LAM (Laboratoire d'Astrophysique de Marseille) UMR7326, 13388, Marseille, France
\and
ESAC, ESA, PO Box 78, Villanueva de la Cañada, Madrid 28691, Spain
\and
Infrared Processing and Analysis Center, California Institute of
Technology, Pasadena, CA 91125, U.S.A.
\and
European Southern Observatory, Karl-Schwarzschild Str. 2, D-85748 Garching, Germany
\and
European Southern Observatory, ESO Vitacura, Alonso de Cordova 3107, Vitacura,
Casilla 19001, Santiago, Chile
\and
Atacama Large Millimeter/submillimeter Array, ALMA Santiago Central Offices,
Alonso de Cordova 3107, Vitacura, Casilla 763-0355, Santiago, Chile
\and
School of Earth and Space Exploration, Arizona State University, Tempe, AZ 85287, USA
\and
UPMC Univ Paris 06, UMR7095, Institut d'Astrophysique de Paris, 75014, Paris, France
\and
Department of Physics, University of Oxford, Denys Wilkinson Building,
Keble Road, Oxford OX1 3RH, UK
} 
\titlerunning{\Planck's Dusty GEMS}
\authorrunning{Ca{\~n}ameras et al.}
\date{Received / Accepted }

\abstract{
We present an analysis of CO spectroscopy and infrared-to-millimetre
dust photometry of 11 bright far-infrared/submillimetre
sources discovered through a combination of the \Planck\ all-sky
survey and follow-up \textit{Herschel}-SPIRE imaging -- ``\Planck's
Dusty Gravitationally Enhanced subMillimetre Sources.''  Each source
has a spectroscopic redshift $z=2.2$--3.6 from 
a blind redshift search with EMIR at the IRAM 30-m
telescope. Interferometry obtained at IRAM and the SMA, and
optical/near-infrared imaging obtained at the CFHT and the VLT
 reveal morphologies consistent with strongly gravitationally
lensed sources. Additional photometry
was obtained with JCMT/SCUBA-2 and IRAM/GISMO at 850$\,\mu$m and
2\,mm, respectively. The spectral energy distributions of our sources
peak near either the 350$\,\mu$m or 500$\,\mu$m bands of SPIRE. 
All objects are bright, isolated point sources in the 18\arcsec\ beam of
SPIRE at 250$\,\mu$m, with apparent far-infrared luminosities of up to
$3\times10^{14}\,{\rm L}_{\odot}$ (not correcting for the lensing
effect). Their morphologies and sizes, CO line widths and
luminosities, dust temperatures, and far-infrared luminosities provide
 additional empirical evidence that these are 
strongly gravitationally lensed high-redshift galaxies on the submm
sky.  We
discuss their dust masses and temperatures, and use additional
\textit{WISE\/} 22-$\mu$m photometry and template fitting to rule out
a significant contribution of AGN heating to the total infrared
luminosity. Six sources are detected in FIRST at 1.4\,GHz.
Four have flux densities brighter than expected from the local far-infrared-radio
  correlation, but in the range previously found for high-$z$ submm
  galaxies, one has a deficit of FIR emission, and 6 are consistent
  with the local correlation, although this includes 3 galaxies with
  upper limits. The
global dust-to-gas ratios and star-formation efficiencies of our
sources are predominantly in the range expected from massive,
metal-rich, intense, high-redshift starbursts. An extensive
multi-wavelength follow-up programme is being carried out to further
characterize these sources and the intense star-formation within
them.}


\maketitle
\section{Introduction}
\label{sec:introduction}

The brightest, most strongly gravitationally lensed galaxies in the
high-redshift Universe have an extraordinary potential to advance our
understanding of the processes that regulate the growth of the most
massive galaxies we see today. In particularly fortuitous cases,
gravitational lensing from massive, intervening galaxies or galaxy
clusters not only boosts the apparent integrated brightness of
high-redshift galaxies by factors up to 20--60, but also magnifies
their images by similar factors while conserving surface
  brightness). Thereby, they allow us to study the fine spatial
details of intensely star-forming high-redshift galaxies at scales
much below 1\,kpc, down to around 100\,pc \citep[e.g.,][]{swinbank10,
  swinbank11}. This is more akin to the scales of individual
star-forming regions in nearby galaxies than the galaxy-wide scales
(of order a few kpc) with which we must otherwise content ourselves at
cosmological distances.

Since the discovery of the first gravitationally lensed galaxy in the
optical \citep[][the first gravitationally lensed quasar had already
 been discovered by \citealt{walsh79}]{soucail87}, strongly
gravitationally lensed galaxies have been identified and studied in
all wavebands from the ground and in space. Strongly gravitationally lensed
submm galaxies provide an extraordinary possibility to probe individual
star-forming regions in the most intensely star-forming high-redshift
galaxies.  They are also very promising sources
to increase our understanding of how the
deep gravitational potential well of high-redshift galaxies, their high gas
fractions and gas-mass surface densities, and feedback from star
formation and perhaps active galactic nuclei, are setting the stage for
the intense star-formation at fine spatial detail \citep[e.g.,][]{danielson11, swinbank11, combes12}.

Massive, dust-enshrouded, and
relatively evolved high-redshift galaxies are characterized by the
bright thermal infrared emission from dust heated by intense star
formation during their rapid primordial growth phase. These galaxies
have typical stellar and dynamical masses of a few times
$10^{10-11}\,{\rm M}_\odot$ \citep[][]{smail04, swinbank06, nesvadba07}
and form stars at prodigious rates of up to about
$1000\,{\rm M}_\odot\,{\rm yr}^{-1}$, which
are unparallelled in the nearby Universe. As a population, dusty
starburst galaxies may have contributed as much as about half of the
total energy production from star formation at these cosmic epochs
\citep[e.g.,][]{hauser01,dole06}.

Unfortunately, given their importance for our understanding of
high-redshift star formation and galaxy growth, the large stellar and
dynamical masses and short evolutionary timescales of these galaxies
make them very rare on the sky. The densities of far-infrared and
submillimetre selected (FIR/submm) galaxies that are bright enough to
be good candidates for strong gravitational lensing are only around
one every few square degrees for sources with $S_{\rm
  500}\approx100$\,mJy, with large uncertainties. For example,
\citet{lapi12} predict about 0.003--0.1$\,{\rm deg}^{-2}$ for sources
with $S_{\rm 500}\ge400\,$mJy on the sky, for expected maximal
gravitational magnification factors of 20--30 adopted in the
  models. Consequently, we may expect to find only a few of these
sources on the entire extragalactic sky, in accordance with models of
the diffuse infrared background light
\citep[][]{bethermin12}. Nonetheless, at luminosities above about
$10^{13}{\rm L}_{\odot}$ (corresponding to $S_{\rm 500}
\approx100$\,mJy), they are expected to dominate the integrated
FIR/submm luminosity function \citep[][]{negrello07}.

A major breakthrough has been the recent discovery by
\textit{Herschel\/}
\footnote{\textit{Herschel\/} is an ESA space observatory with science
  instruments provided by European-led Principal Investigator
  consortia and with important participation from NASA.}  and the
South-Pole Telescope (SPT) of sizeable sets of strongly
gravitationally lensed submillimetre galaxies, with typical
magnification factors of a few. These are using the new
generation of wide-field surveys (several thousands of square
degrees), probing in particular the range in FIR/submm flux density
between 100 and 200\,mJy 
  \citep[e.g.,][]{negrello10,harris12,vieira13,wardlow13,bussmann13}.
Most gravitationally lensed sources identified in these surveys are
magnified by individual massive galaxies at intermediate redshifts,
producing partial Einstein rings of a few arcsec in diameter and
magnification factors up to about 10. However, the number of sources
identified in these surveys above flux densities of 200--300\,mJy
remains rather small. For example, \citet{bussmann13} list 30
  sources with far-infrared fluxes in this range, with 6 sources above
  $S_{\rm 350}=$350~mJy. So far, to our knowledge only one source with
$S_{\rm 350}>500$\,mJy at 350$\,\mu$m has been published by the SPT
collaboration \citep[][]{vieira13}. Meanwhile, the brightest source in
the \citet{bussmann13} sample has $S_{\rm 350}=$484~mJy.

The extragalactic \textit{Herschel\/} and SPT surveys together cover
about 4\,\% of the sky, which highlights the importance of having an
all-sky survey to identify in a systematic way high-redshift FIR/submm
galaxies above the 100--300\,mJy regime. These sources are likely to
be amongst the most strongly gravitationally lensed galaxies on the
sky.  \Planck\
is the first all-sky survey in the submillimetre with
the depth and spatial resolution necessary to probe the brightest, and
presumably most strongly gravitationally lensed high-redshift infrared
galaxies observable to us. The 90\,\% completeness limit of the \Planck\
Catalogue of Compact Sources (PCCS) corresponds to
$L_{\rm IR}\approx 6\times 10^{13} {\rm L}_\odot$
at $z=2$ \citep[][]{planck2013-p05}. At these luminosities, even all-sky
surveys may reveal only very small numbers of sources
\citep[][]{negrello10,bethermin12,lapi12}. For example,
\citet{herranz13} find only very small numbers of high-redshift
sources in the Early Release Catalogue of Compact Sources (ERCSC) from
Planck (\Planck\ Collaboration et al. 2011), primarily
blazars; most extragalactic sources in the ERCSC are low-redshift
galaxies.

We used photometry derived from \Planck-HFI \citep[described
  in][]{planck2013-p01,planck2013-p03} to identify all compact sources
in the \Planck\ maps that have colours consistent with being
exceptionally bright, dusty, intensely star-forming high-redshift
galaxies.  We then obtained far-infrared photometry of the most
  promising candidates \citep[the ``HPASSS'' programme][Paper~I
    hereafter]{planck2014-XXVII}, using the Spectral and Photometric
  Imaging Receiver (SPIRE) on board the \textit{Herschel\/} space
    telescope. This sample, obtained through \textit{Herschel\/}
``Must-Do'' Director's Discretionary Time, includes 234 of
the brightest, rarest sources in the sky (one source per several tens
of square degrees, Planck Collaboration, 2015, in prep.). The sample
was deliberately selected to only include sources that do not fall
into the large {\it Herschel\/} survey fields, and which are new to
the literature. SPIRE confirms that these sources have the typical
submm colours of high-$z$ infrared and submm galaxies. Most of the 234
sources of the HPASSS sample are overdensities of multiple galaxies
with the typical FIR colours of high-$z$ galaxies, and only four are
Galactic cirrus clouds. Another small subset are bright individual,
isolated point sources in the 20\arcsec\ beam of SPIRE, consistent
with being exceptionally bright, presumably strongly gravitationally
lensed high-redshift galaxies.  An overview of the HPASSS sample is
given in Paper~I.

Here we present the first results of our multi-wavelength follow-up of
the 11 brightest of these isolated HPASSS sources (``\Planck's
Dusty GEMS''), which can be observed from the northern hemisphere.
All have flux densities at 350$\,\mu$m measured with SPIRE that are at
least $S_{350}\approx 300$\,mJy, well above the typical range probed by
the SPT and wide-field {\it Herschel\/} surveys. The brightest source has
$S_{350}=1050$\,mJy.  Welikala et al. (2015, A\&A submitted) discuss another
set of weaker gravitational lenses observed with \Planck\ and the SPT;
however, the 11 sources we discuss here were previously unknown and
are substantially brighter. Another source, PLCKERC857
PLCK\_G270.59$+$58.52, taken from the ERCSC, which also satisfies our
selection criteria and fortuitously falls into the H-ATLAS survey
area, has already been discussed by \citet{fu12}, \citet{herranz13},
and \citet{bussmann13}. HLS~091828.6$+$514223 at $z=5.2$
\citep{combes12,rawle13} behind the galaxy cluster Abell~773, has been
discovered independently from our survey as part of the
\textit{Herschel\/} Lensing Survey \citep{egami10}, but is also
included in our \Planck\ parent sample (Paper~I). It forms a
1.5\arcsec, near-complete Einstein ring with a magnification factor of
$9\pm2$ \citep{rawle13}.

Our present analysis has three main goals. Firstly, we determine
spectroscopic redshifts and provide empirical evidence from millimetre
photometry and spectroscopy that these are indeed high-redshift
sources, which owe their exceptional brightness in the submm to strong
gravitational lensing. We also show ground-based near-infrared imaging,
illustrating that the submm sources lie behind overdense regions in the
intermediate-redshift Universe. Secondly, we characterize their global
dust and star-formation properties and show that their observed
far-infrared emission is dominated by star formation, not by powerful
obscured quasars. Thirdly, we estimate molecular gas masses from their
CO line emission, calculate dust-to-gas mass ratios, and use their FIR
luminosities and molecular gas mass estimates to show that they
are more akin to very rapidly forming ``starburst'' galaxies, than the 
more gradually, but still intensely star-forming high-redshift
galaxies on the ``main sequence'' \citep[e.g.,][]{elbaz11}. This is the
first in a series of papers about ``Planck's Dusty GEMS,'' and for
this analysis we only use parts of the comprehensive data sets we
already have in hand.  Subsequent publications will provide lensing
model solutions, discuss the spatially-resolved properties of our
sources in a number of wavebands, and use the multiple millimetre line
detections to investigate the detailed gas and star-formation
properties of these systems.
\begin{figure*}
\centering
\includegraphics[width=0.8\textwidth]{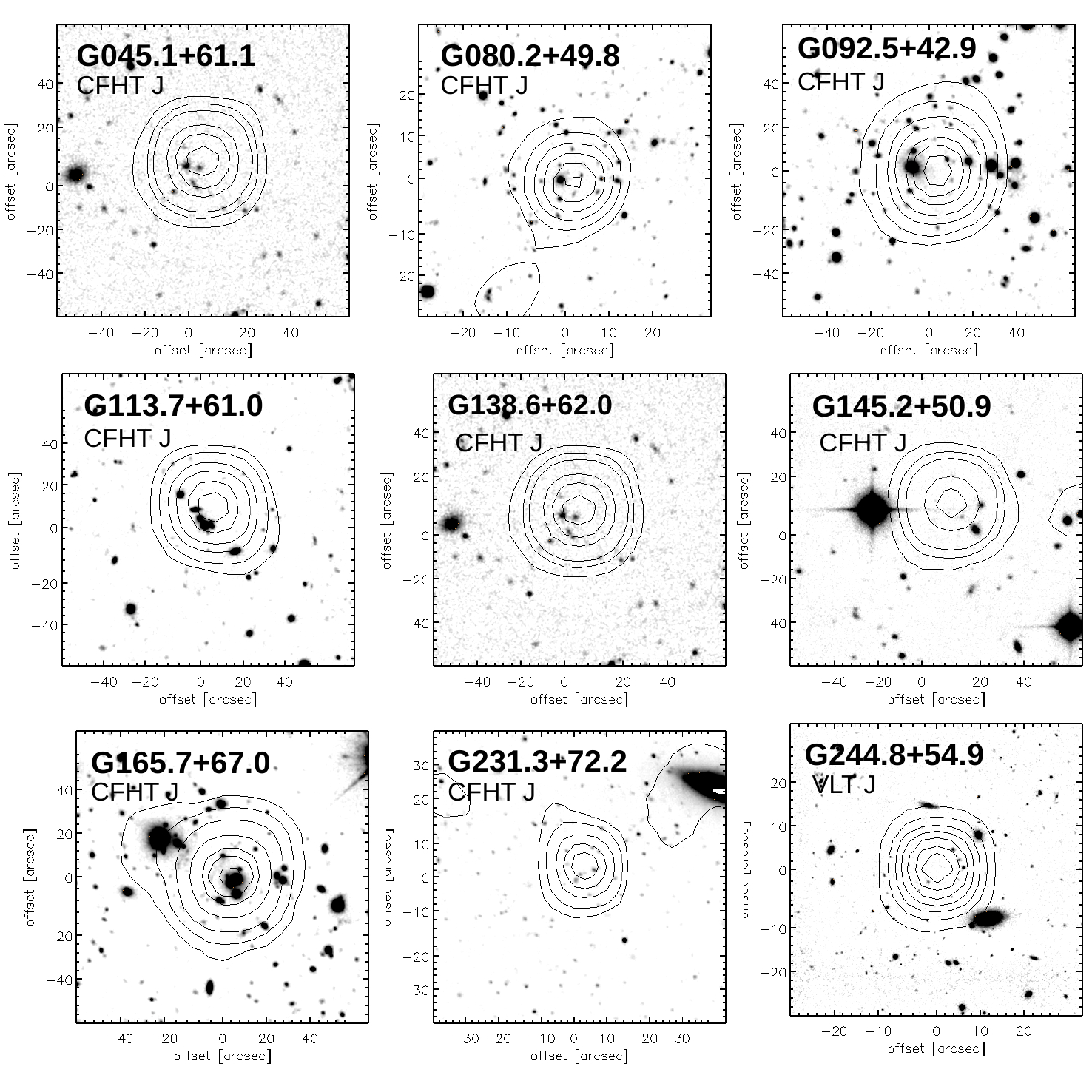}
\caption{Near-infrared $J$-band images of the nine sources for which
  we already have optical/NIR imaging and millimetre interferometry
  (Fig.~\ref{fig:morphsma}).  Contours show the {\it Herschel}-SPIRE
  morphology at 250$\,\mu$m.} 
\label{fig:morphologies}
\end{figure*}

The outline of the paper is as follows.  After the descriptions of our
sample selection in Sect.~\ref{sec:selection}, and our follow-up
observations and data reduction in Sect.~\ref{sec:observations}, we
provide the redshifts in Sect.~\ref{sec:spectra}, and the FIR
luminosities, dust masses and temperatures in Sect.~\ref{sec:bbfit}. In
Sect.~\ref{sec:lensing} we present the evidence that our sources are very
strongly gravitationally lensed galaxies, including arcsec-resolution
submm and millimetre interferometry of the dust emission, as well as
NIR/optical imaging. In Sect.~\ref{sec:templatefit} we impose additional
constraints from the \textit{Wide-field Infrared Survey Explorer,
 WISE\/} and the VLA FIRST survey to demonstrate that the
  far-infrared spectral energy distributions of our sources are
  not dominated by radiation from powerful AGN, before turning to
their gas and star-formation properties in
Sect.~\ref{sec:dustgasratios} and \ref{sec:SK}, respectively. We
summarize our results in Sect.~\ref{sec:summary}. Throughout the paper
we adopt a flat $H_0= 70\,{\rm km}\,{\rm s}^{-1}\,{\rm Mpc}^{-1}$
concordance cosmology with $\Omega_{\rm M}=0.3$ and
$\Omega_{\Lambda}=0.7$.

\section{Sample selection}
\label{sec:selection}

In Fig.~\ref{fig:morphologies} we show 9 of our 11 newly
  discovered gravitational lens candidates, which are the brightest
  amongst our sources selected from the nominal data release of the
  \Planck\ all-sky survey. The parent sample consists of two subsets
that were identified with similar colour cuts in the 857-GHz to
545-GHz flux density ratio of $S_{857}/S_{545} < 1.5$--2.  This is
well-matched to the expected spectral shape of the far-infrared
continuum of dusty starburst galaxies at redshifts $z\ga2$. Six of our
sources were taken from the PCCS, which includes all sources with
${\rm S/N}>4$ at 545\,GHz on the cleanest 52\,\% of the sky
\citep[][]{planck2013-p05}.  Five sources come from a dedicated, blind
search for high-redshift candidates in the \Planck\ maps, which probes
fainter sources in the cleanest 35\,\% of the sky after subtracting
estimates of the cosmic microwave background and Galactic cirrus
emission. Table~\ref{tab:photometry} lists the origin of each
target. The second subsample will be described in detail by Planck
Collaboration 2015 (in prep.).  A comprehensive summary of the
selection and in particular the cleaning algorithm adopted to identify
this subsample is given in Paper~I.

\begin{figure*}
\centering
\includegraphics[width=0.8\textwidth]{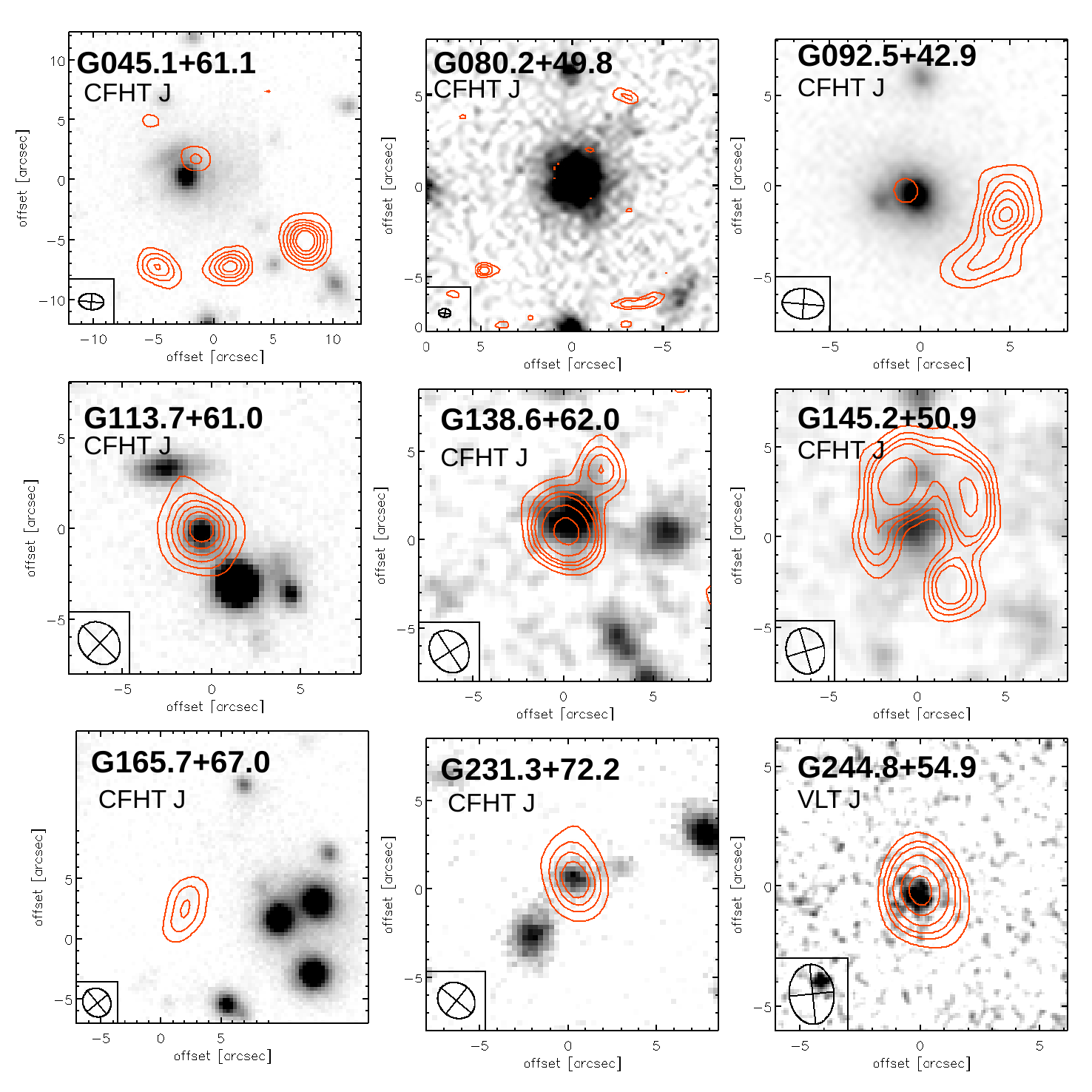}
\caption{Near-infrared $J$-band images of the nine sources from
  Fig.~\ref{fig:morphologies}.  Contours show the SMA morphology at
  850$\mu$m. The FWHM beam size of all SMA data is about
  2\arcsec\ (the FWHM beam size is shown in the bottom left corner of
  each panel), except for PLCK\_G080.2$+$49.8, where it is 0.8\arcsec.}
\label{fig:morphsma}
\end{figure*}

\begin{table*}[htbp]
\begingroup
\newdimen\tblskip \tblskip=5pt
\caption{Mid-infrared to millimetre photometry.
The columns are: source name; indication of whether the sources come from the
PCCS or our own high-$z$ selection (OT2, see Sect.~\ref{sec:selection});
RA and Dec coordinates; redshift of the high-z source;
SDSS redshift of the intervening foreground source;
\textit{WISE\/} 22-$\mu$m flux density;
\textit{Herschel}-SPIRE 250-$\mu$m flux density;
\textit{Herschel}-SPIRE 350-$\mu$m flux density;
\textit{Herschel}-SPIRE 500-$\mu$m flux density;
JCMT/SCUBA-2 850-$\mu$m flux density; and
IRAM 30-m/GISMO 2-mm flux density.}
\label{tab:photometry}
\vskip -5mm
\footnotesize
\setbox\tablebox=\vbox{
 \newdimen\digitwidth
 \setbox0=\hbox{\rm 0}
 \digitwidth=\wd0
 \catcode`*=\active
 \def*{\kern\digitwidth}
 \newdimen\signwidth
 \setbox0=\hbox{+}
 \signwidth=\wd0
 \catcode`!=\active
 \def!{\kern\signwidth}
 \newdimen\pointwidth
 \setbox0=\hbox{.}
 \pointwidth=\wd0
 \catcode`@=\active
 \def@{\kern\pointwidth}
 \halign{\tabskip=0pt\hbox to 1.3in{#\leaderfil}\tabskip=1em&
 \hfil#\hfil\tabskip=1em&
 \hfil#\hfil\tabskip=1em&
 \hfil#\hfil\tabskip=1em&
 \hfil#\hfil\tabskip=0.5em&
 \hfil#\hfil\tabskip=1em&
 \hfil#\hfil\tabskip=1em&
 \hfil#\hfil\tabskip=1em&
 \hfil#\hfil\tabskip=1em&
 \hfil#\hfil\tabskip=1em&
 \hfil#\hfil\tabskip=1em&
 \hfil#\hfil\tabskip=0pt\cr
\noalign{\doubleline}
\omit\hfil Source\hfil& Survey& RA& Dec& $z_{\rm source}$& $z_{\rm fg,sdss}$ & $S_{22}$& $S_{250}$& $S_{350}$& $S_{500}$& $S_{850}$& $S_{2000}$\cr
\noalign{\vskip 3pt}
\omit& & (J2000)& (J2000)& & & [mJy]& [mJy]& [mJy]& [mJy]& [mJy]& [mJy]\cr
\noalign{\vskip 4pt\hrule\vskip 4pt}
PLCK\_G045.1+61.1&  OT2& 15:02:36.04& +29:20:51& 3.4*& 0.56$^{\rm \,a,b}$*& *$<$5.4 & *161$\pm$*5& *328$\pm$*6& 397$\pm$*6& 130$\pm$13& 13@*$\pm$2@*\cr
PLCK\_G080.2+49.8&  OT2& 15:44:32.40& +50:23:46& 2.6*& 0.5*  & *$<$5.4& *220$\pm$*9& *340$\pm$*6& 314$\pm$*6& *87$\pm$*5& *7@*$\pm$2@*\cr
PLCK\_G092.5+42.9& PCCS& 16:09:17.76& +60:45:21& 3.3*& 0.5*  & *$<$5.4& *765$\pm$*7& *865$\pm$*8& 696$\pm$*7& 191$\pm$16& *8.4$\pm$0.6\cr
PLCK\_G102.1+53.6&  OT2& 14:29:17.98& +59:21:09& 2.9*& 0.74*$^{\rm \,a}$ & 8.1$\pm$0.7& *327$\pm$*4& *410$\pm$*5& 339$\pm$*5& \dots& *5.5$\pm$0.7\cr
PLCK\_G113.7+61.0& PCCS& 13:23:02.88& +55:36:01& 2.4*& 0.3*  & *6.0$\pm$0.7& *672$\pm$*7& *603$\pm$*6& 373$\pm$*5& 108$\pm$10& *4@$*\pm$1@*\cr
PLCK\_G138.6+62.0& PCCS& 12:02:07.68& +53:34:40& 2.4*& 0.4*  & 6.2$\pm$0.8& *619$\pm$*6& *664$\pm$*8& 474$\pm$*6& 123$\pm$14& *7@$*\pm$1@*\cr
PLCK\_G145.2+50.9& PCCS& 10:53:22.56& +60:51:49& 3.6*& \dots & *$<$5.4& *453$\pm$*5& *719$\pm$*7& 781$\pm$*8& 360$\pm$23& 38@$*\pm$4@*\cr
PLCK\_G165.7+67.0& PCCS& 11:27:14.60& +42:28:25& 2.2*& 0.34*$^{\rm \,a,b}$ & 10.4$\pm$0.9& *867$\pm$*8& *753$\pm$*6& 472$\pm$*5& *90$\pm$*9& *8@$*\pm$1@*\cr
PLCK\_G200.6+46.1&  OT2& 09:32:23.67& +27:25:00& 3.0*& 0.6* & *$<$5.4 & *209$\pm$*4& *294$\pm$*4& 273$\pm$*5& 110$\pm$10 & *7@$*\pm$2@*\cr
PLCK\_G231.3+72.2&  OT2& 11:39:21.60& +20:24:53& 2.9*& 0.1* & *$<$5.4 & *299$\pm$*4& *401$\pm$*5& 341$\pm$*6& 111$\pm$12& *9@*$\pm$1@*\cr
PLCK\_G244.8+54.9& PCCS& 10:53:53.04& +05:56:21& 3.0*& 0.13* & *$<$5.4 & 1050$\pm$10& 1054$\pm$10& 777$\pm$*7& 198$\pm$11& 19@*$\pm$2@*\cr
\noalign{\vskip 4pt\hrule\vskip 6pt}
HLS-J0918$^{\rm \,c}$&    & 09:18:28.6&   +51:42:23& 5.2*& & & **85$\pm$*8& *168$\pm$*8& 203$\pm$*9& 125$\pm$*8& 15@*$\pm$7@*\cr
HATLAS-J1146$^{\rm \,d}$& & 11:46:37.9& $-$00:11:32& 3.3& & & *323$\pm$24& *378$\pm$28& 298$\pm$24& *93$\pm$12& 38@*$\pm$6@*\cr
\noalign{\vskip 4pt\hrule\vskip 6pt}
}}
\endPlancktablewide
\tablenote{{\rm a}} Spectroscopic redshift of intervening foreground source from the SDSS.\par
\tablenote{{\rm b}} Multiple foreground sources at a common spectroscopic redshift near the line of sight.\par
\tablenote {{\rm c}} This source has been found with the
 \textit{Herschel\/} Lensing Survey and has previously been discussed by
 \citet{combes12} and \citet{rawle13}. The last column lists the 1.2-mm flux
 density measured with MAMBO on the IRAM 30-m telescope.\par
\tablenote {{\rm d}} This source from the ERCSC \citep{planck11} falls
 serendipitously into the HATLAS field and has been previously
 discussed by \citet{fu12}. The last column lists the 1.2-mm flux density
 measured with MAMBO on the IRAM 30-m telescope.\par
\endgroup
\end{table*}

All sources were followed up with \textit{Herschel}-SPIRE photometry
at 250$\,\mu$m, 350$\,\mu$m, and 500$\,\mu$m as part of the HPASSS
survey, mostly during ``Must-Do'' Director's Discretionary Time, and
is a subset of the sample presented in Paper~I.  SPIRE has about 10
times greater depth and 20 times higher spatial resolution than
\Planck, with beam FWHMs of 18\arcsec, 24\arcsec, and
35\arcsec\ at 250$\,\mu$m, 350$\,\mu$m, and 500$\,\mu$m,
respectively\footnote{See the SPIRE Handbook 2014,
\url{http://herschel.esac.esa.int/Docs/SPIRE/spire_handbook.pdf}.}.
Therefore, SPIRE can be used to identify and localize
the counterparts of \Planck\ sources in the same wavelength
regime. We identified all potential gravitationally lensed galaxies as
bright, isolated point sources in the SPIRE images, with spectral
energy distributions peaking at flux densities above 300\,mJy in either
the 350$\,\mu$m or the 500$\,\mu$m band as observed with SPIRE (in the
final calibration of our SPIRE maps, one source has
$S_{350}$=294\,mJy).  This is well matched to the 90\,\% completeness
limit of the PCCS of about 600\,mJy \citep{planck2013-p05}, and
purposefully excludes similar sources already identified in other
surveys.

We find a total of 15 gravitational lens candidates. For the purposes
of this paper, we exclude the two already described in the literature
\citep[][see Sect.~\ref{sec:introduction}]{fu12, combes12}, and another
two that are in the far South. The source described by \citet{combes12} is at
a higher redshift ($z=5.2$) than the sources we discuss here. Apart from
that, including those other sources would not change our general
conclusions. 

In the present analysis we discuss the 11 Northern
sources that are already spectroscopically confirmed to be at high
redshifts. It is worth noting that by focusing on single, very bright
sources in the Herschel images, we may have missed sets of multiple, but fainter
gravitationally lensed objects behind the same intervening structures;
these we would have identified as overdensities of high-redshift
infrared galaxies in the overall HPASSS sample (Paper~I). Irrespective
of this, given the extraordinary sky coverage of our parent sample and
the brightness of our targets, six of which reach or even exceed the
completeness limits of the PCCS in the 353, 545, and 857\,GHz bands of
330, 570, and 680\,mJy, respectively \citep[Table~1 of][]{planck2013-p05},
the 11 sources discussed here are likely to be amongst the brightest
individual high-redshift galaxies on the sky.

\section{Photometry}
\label{sec:observations}

\subsection{\textit{Herschel}-SPIRE FIR photometry}
We base our analysis on the \textit{Herschel}-SPIRE photometry at 250$\,\mu$m,
350$\,\mu$m, and 500$\,\mu$m, as well as on ground-based JCMT/SCUBA-2
850-$\mu$m and IRAM/GISMO 2-mm single-dish photometry.  For sources
in the range $z=2.2$--3.6, this corresponds to a wavelength range 
50--80$\,\mu$m and 400--600$\,\mu$m in the rest-frame, 
which cover the expected peak and Rayleigh-Jeans tail, respectively, of the
warm dust emission heated by intense star formation.

\textit{Herschel}-SPIRE observations were carried out between December~2012 and
March~2013. The \textit{Herschel}-SPIRE photometry was obtained using
{\tt STARFINDER}
\citep[][]{diolaiti00} as part of the HPASSS survey, and is discussed
in detail in Paper~I. {\tt STARFINDER} integrates over the point
spread function obtained directly from the image, and is therefore
more suitable for data dominated by confusion noise than classical
aperture photometry. Measured flux densities are between 294\,mJy and 1054\,mJy
at 350$\,\mu$m and between 270\,mJy and 800\,mJy at 500$\,\mu$m,
respectively (Table~\ref{tab:photometry}). Typical uncertainties are
of the order of 5--10\,mJy, which is close to the confusion level in
the maps \citep[][]{nguyen10}, but only includes the measurement
error, not the systematic uncertainty of 7\,\% inherent in the SPIRE
photometry.  However, this has no impact on our results.

\subsection{JCMT/SCUBA-2 photometry at 850$\,\mu$m}
The SCUBA-2 \citep{scuba2} data were taken between September 2012 and
May~2014 in moderate conditions with individual observing times
  of 15~min per source. The data were reduced using a configuration
file optimized for point-source calibrators using the {\tt smurf} data
reduction software package for SCUBA-2 \citep{smurf}.  Flux densities
were extracted using aperture photometry with a 30\arcsec\ diameter
aperture, where the background was estimated within an annulus with
inner and outer diameters of 37.5\arcsec and 60\arcsec,
respectively. Uncertainties are between 4~mJy and 21~mJy per beam.
These flux densities are then corrected for missing flux density due
to the aperture size by dividing by 0.85, as described by
\cite{dempsey2013}. An 8\,\% calibration uncertainty is added in
quadrature to the photometric errors. The resulting 850$\,\mu$m flux
densities are listed in Table~\ref{tab:photometry}.

\subsection{SMA 850$\,\mu$m interferometry}
All sources were also observed in the continuum with the SMA in the
850$\,\mu$m band. Observations were carried out between June~2013 and
June~2014.  Data were taken under good to excellent conditions 
  with pwv$<$2~mm in shared-track mode to obtain good {\it uv}
coverage, in spite of observing each source with less than one track;
this was made possible by their extraordinary brightness in the
submm. Integration times per source are between 2 and 7~hrs. All
sources but PLCK\_G080.2$+$49.8 were observed through programme
2013B-S050 in the compact configuration, with a beam of about
$2\arcsec\times2\arcsec$. PLCK\_G080.2$+$49.8 was observed through DDT
programme 2013A-S075 in the extended configuration, giving a beam of
$0.8\arcsec\times0.5\arcsec$. Data were calibrated in IDL using the
{\tt MIR} package, and analysis and imaging utilized the {\tt MIRIAD}
package.  A full discussion of the interferometry is beyond the scope
of the present analysis, and will be presented in a subsequent
publication. Here we only use the 850$\,\mu$m morphologies to
illustrate that these are indeed strongly gravitationally lensed
galaxies. Comparison with the single-dish flux densities from SCUBA-2
suggests that we recover at least 80--90\,\% of the total flux
density, implying that we have typically not missed fainter, more
extended components.

\subsection{IRAM-30-m /GISMO 2-mm photometry}

Two-mm continuum observations were carried out with the 30-m telescope
of IRAM between 17 and 23 April with 1.3-1.7~mm of precipitable water
vapor, and between 29 October and 5 November, 2013, with 3.4-8.6~mm of
preciptable water vapor (programmes 222-12 and 100-13).

We
used the $8\times16$ pixel bolometer camera GISMO, which covers a 
frequency range of 140--162\,GHz \citep[][]{staguhn12} in the
2-mm band. At the 30-m telescope of IRAM, GISMO covers 
a $1.8\arcmin\times3.7\arcmin$ field of view with a
$21\arcsec\times21\arcsec$ beam. We used
$2\arcmin\times2\arcmin$ Lissajous maps with a relative flux
density stability of about 8\,\%, which is optimized to obtain high
signal-to-noise ratios of relatively faint objects like ours, with only a
small number of bad channels. Total integration times per source
were between 10 and 100 minutes, depending on the expected flux density of the 
target.

We used the {\tt CRUSH} software package
\citep[][]{kovacs13} to reduce and calibrate individual scans, and to
combine them into the final image.  We used the ``faint'' option, which
is well adapted to signals around 10\,mJy, detected at signal-to-noise
ratios of less than 10 per scan. We detected all targets as point sources, 
with S/N between 3.4 and 14.8.

\subsection{CFHT and VLT optical/NIR photometry}

In order to identify and broadly characterize the foreground lensing
structure, we also obtained near-infrared optical imaging of our
sample using Megacam and WIRCAM on the Canada-France-Hawaii
Telescope (CFHT), and with HAWKI and FORS on the Very Large Telescope
(VLT) of the European Southern Observatory (ESO). Here we use the
optical imaging to highlight that our sources lie behind massive
intervening structures, either galaxy groups or clusters. Observations
of the overall sample are still on-going, but have already provided us
with optical or NIR imaging of all but one target in at least one
band. We used the {\tt scamp} and {\tt swarp} software
 \citep[][]{bertin10b,bertin10a} to register our images relative to
the USNO-B2 catalog, with a typical positional uncertainty of
0.2\arcsec$-$0.5\arcsec. A full discussion of these data and what
they imply more quantitatively for the lensing configuration will be
presented in a subsequent publication.


\section{Blind spectroscopic redshift survey in the millimetre}

\subsection{IRAM 30m/EMIR spectroscopy}

We performed a blind redshift search in the 3-mm and 2-mm bands for
all 11 targets using the wide-band heterodyne receiver EMIR at the 30-m
telescope of IRAM. Following a pilot programme to measure a
spectroscopic redshift of our first source, PLCK\_G080.2$+$49.8, through
regular programme 82-12 and Director's Discretionary programme D05-12, we
obtained another 75~hrs of observing time through Director's Discretionary
programme D09-12 and the regular programme 094-13 in April and June
2013. For all sources we used the WILMA and FTS backends during good
to variable conditions.

Individual scans were 30 seconds long, and we observed sets of 12
scans followed by a calibration. Data were reduced using {\tt CLASS}
\citep[][]{gildas13}.  We took advantage of a dedicated routine kindly
provided by C. Kramer to individually correct the baselines in each of the
sub-bands of the FTS backend. The full set of lines will
be discussed in a subsequent publication; here we only use the
lowest-$J$ CO transition available in each source, which
provides additional empirical evidence that these are strongly gravitationally
lensed galaxies. The redshifts are also derived from these lines, and
found to be consistent with the full set of available lines per galaxy.

\subsection{Spectroscopic redshifts}
\label{sec:spectra}

To obtain a spectroscopic redshift for each source, we started with a
blind line search in the 3-mm atmospheric window, which we cover
almost entirely with two interleaved tunings of EMIR centred at
89.4\,GHz and 97.4\,GHz, respectively. We used the WILMA and FTS
backends in parallel, which have band widths of about 4\,GHz and 8\,GHz,
respectively.

\begin{figure*}
\centering
\includegraphics[width=0.95\textwidth]{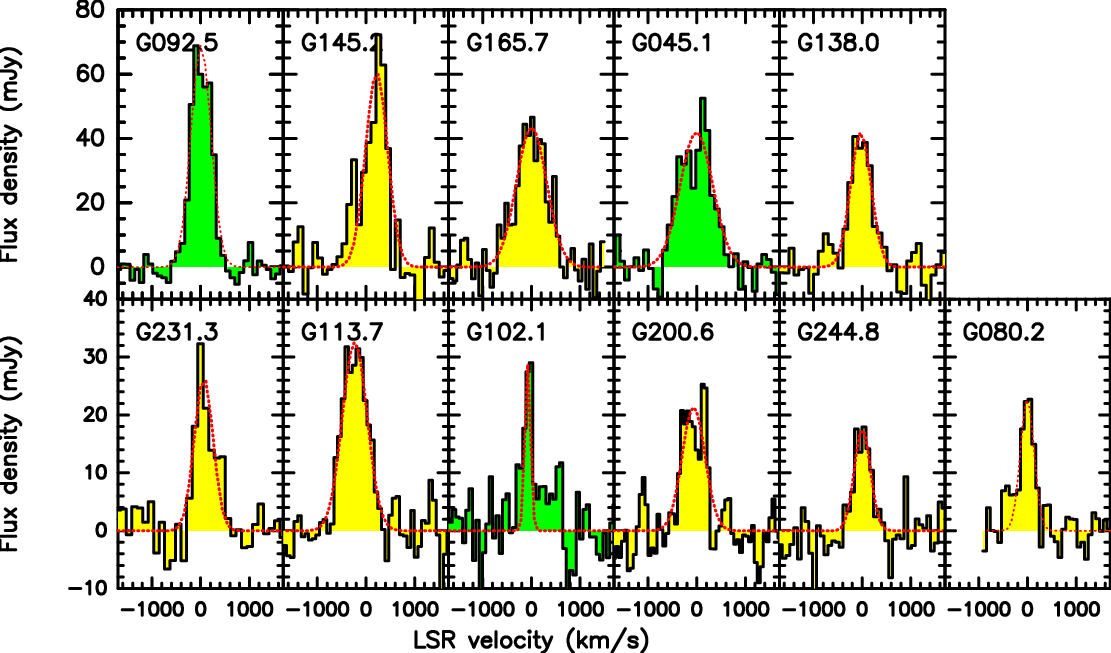}
\caption{Lowest-$J$ CO emission line in each of our targets.
  CO(3-2) and CO(4-3)
  are indicated in yellow and green, respectively. 
  Red dotted lines show simple
  single-Gaussian fits.}
\label{fig:lines}
\end{figure*}

We discovered a bright emission line in each source in one of the
tunings in the 3-mm band (Fig.~\ref{fig:lines}). Subsequently we
calculated all possible redshifts compatible with the observed
frequency of the line, and tested our redshift hypothesis by searching
for a second line (typically CO(5-4) or
CO(4-3)) at the predicted
higher frequency. This required a separate frequency tuning, with the
choice depending on each redshift hypothesis (typically lying in the 2-mm
band). We started with the redshift that was closest to the
photometric redshift derived from our SPIRE photometry, assuming a
dust temperature of $T_{\rm d}=30$\,K, which is the temperature of the
dust component that dominates the FIR spectral energy distribution of
the Cosmic Eyelash \citep[][]{ivison10}. This yielded the correct
redshift in all but the two galaxies, specifically those with the highest dust
temperatures (Table~\ref{tab:dust}).  In the end, our EMIR follow-up
spectroscopy led to accurate and secure spectroscopic redshifts for
all 11 targets, with 2--8 lines detected per source; this provides a
wide range of constraints on the physical properties of the gas in
these galaxies, which we will discuss in more detail in forthcoming
papers.

\begin{table*}[htbp]
\begingroup
\newdimen\tblskip \tblskip=5pt
\caption{Fitted parameters to the FIR dust continuum and dust
    properties.  The columns are: best-fit wavelength of the peak of
    the blackbody in the FIR; peak flux density at $\lambda_{\rm max}$
    of the blackbody; FIR luminosity integrated from
      8-1000$\mu$m, including the magnification factor $\mu$ from our
    modified blackbody fit, and neglecting the possible presence of
    additional, hotter dust components; FIR luminosity derived from
    {\tt DecompIR}, and including the fiducial mid-infrared flux from
    star formation (see
    Sect.~\ref{sec:templatefit}); star-formation rate, including
    magnification factor $\mu$; dust temperature; dust mass,
    including the magnification factor $\mu$; and AGN contribution to the FIR luminosity as obtained with {\tt DecompIR}, where such a component was found.}
\label{tab:dust}
\vskip -3mm
\footnotesize
\setbox\tablebox=\vbox{
 \newdimen\digitwidth
 \setbox0=\hbox{\rm 0}
 \digitwidth=\wd0
 \catcode`*=\active
 \def*{\kern\digitwidth}
 \newdimen\signwidth
 \setbox0=\hbox{+}
 \signwidth=\wd0
 \catcode`!=\active
 \def!{\kern\signwidth}
 \newdimen\pointwidth
 \setbox0=\hbox{.}
 \pointwidth=\wd0
 \catcode`@=\active
 \def@{\kern\pointwidth}
 \halign{\tabskip=0pt\hbox to 1.5in{#\leaderfil}\tabskip=1em&
 \hfil#\hfil\tabskip=1em&
 \hfil#\hfil\tabskip=1em&
 \hfil#\hfil\tabskip=1em&
 \hfil#\hfil\tabskip=1em&
 \hfil#\hfil\tabskip=1em&
 \hfil#\hfil\tabskip=1em&
 \hfil#\hfil\tabskip=1em&
  \hfil#\hfil\tabskip=0pt\cr
\noalign{\doubleline}
\noalign{\vskip -3pt}
\omit\hfil Source\hfil& $\lambda_{\rm max}$& $S_{\rm max}$&
 $\mu\,L_{\rm FIR}$& $\mu\,L_{\rm FIR}^{\rm TPL}$&
 $\mu\,{\rm SFR}$& $T_{\rm d}$&
 $\mu\,M_{\rm d}$ & $L_{\rm FIR,AGN}/L_{\rm FIR,tot}$\cr 
\noalign{\vskip 3pt}
\omit& [$\mu$m]&[ mJy] & [$10^{13} {\rm L}_\odot$]& [$10^{13}{\rm L}_\odot$]&
 [${\rm M}_\odot{\rm yr}^{-1}$]& [K]& [$10^{9}{\rm M}_\odot$] & [\%] \cr
\noalign{\vskip 4pt\hrule\vskip 4pt}
PLCK\_G045.1+61.1& 420$\pm$10& *360$\pm$10& *8.4$\pm$0.1 & 12.1$\pm$0.1 & 14462$\pm$172  & 36.0$\pm$0.5  & *6.5$\pm$0.1 &  \cr
PLCK\_G080.2+49.8& 370$\pm$10& *345$\pm$*5& *4.6$\pm$0.1 & *7.1$\pm$0.1 & *7920$\pm$172  & 33.0$\pm$0.5  & *7.2$\pm$0.2 &  \cr
PLCK\_G092.5+42.9& 322$\pm$*5& *900$\pm$10& 24.8$\pm$0.2 & 34.8$\pm$0.2 & 42698$\pm$344  & 50.1$\pm$0.4  & *5.3$\pm$0.1 & 10 \cr
PLCK\_G102.1+53.6& 333$\pm$*4& *410$\pm$*5& *7.9$\pm$0.1 & 11.9$\pm$0.1 & 13601$\pm$172  & 41.1$\pm$0.3  & *4.3$\pm$0.1 &  \cr
PLCK\_G113.7+61.0& 255$\pm$20& *700$\pm$30& *9.9$\pm$0.2 & 12.6$\pm$0.1 & 17044$\pm$344  & 45.0$\pm$0.4  & *3.5$\pm$0.1 &  \cr
PLCK\_G138.6+62.0& 305$\pm$10& *690$\pm$*5& *9.0$\pm$0.1 & 13.5$\pm$0.1 & 15495$\pm$172  & 38.7$\pm$0.3  & *6.4$\pm$0.1 &  \cr
PLCK\_G145.2+50.9& 400$\pm$10& *785$\pm$30& 21.8$\pm$0.2 & 30.1$\pm$0.2 & 37533$\pm$344  & 40.5$\pm$0.4  & 11.0$\pm$0.1 &  33   \cr
PLCK\_G165.7+67.0& 265$\pm$*2& *875$\pm$*3& 10.3$\pm$0.1 & 13.4$\pm$0.1 & 17733$\pm$171  & 42.5$\pm$0.3  & *5.1$\pm$0.1 &  \cr
PLCK\_G200.6+46.1& 350$\pm$*3& *295$\pm$*2& *5.7$\pm$0.1 & *8.2$\pm$0.1 & *9813$\pm$172  & 37.5$\pm$0.5  & *4.3$\pm$0.1 &  22 \cr
PLCK\_G231.3+72.2& 350$\pm$10& *402$\pm$*5& *7.5$\pm$0.1 & 10.7$\pm$0.1 & 12913 $\pm$172 & 39.3$\pm$0.4  & *4.7$\pm$0.1 &  \cr
PLCK\_G244.8+54.9& 300$\pm$*2& 1135$\pm$*2& 26.5$\pm$0.2 & 36.3$\pm$0.2 & 45625$\pm$344  & 50.0$\pm$0.4  & *5.7$\pm$0.1 & 9 \cr
\noalign{\vskip 4pt\hrule\vskip 4pt}
}}
\endPlancktablewide

\endgroup
\end{table*}

\Planck's Dusty GEMS fall in the redshift range of
$z=2.2$--3.6, comparable to that of radio-selected submillimetre galaxies in
the field \citep{chapman05},
and very similar to that of 
gravitationally lensed submm galaxies in the H-ATLAS survey
\citep[$z=2.1-3.5$;][]{harris12,bussmann13}. The average redshift of
\Planck's Dusty GEMS, $z=2.9\pm0.4$ (we give the width of the
distribution here), is somewhat lower than the redshift range of the bright
gravitationally lensed submm galaxies from the SPT survey.  The SPT sample
has a mean redshift of $z=3.5$, obtained from at least two lines for 12 of
their 26 targets, and a combination of single-line detections and FIR
photometric constraints for the remaining targets \citep[][]{vieira13,weiss13}. 
One reason that we probe somewhat lower
redshifts may be that we select our sources at shorter wavelengths
(350--850$\,\mu$m, compared to 1.4 and 2.0\,mm for the SPT). However,
we stress that our parent sample from \Planck\ does include sources
with higher redshifts, such as HLS~091828.6$+$514223 at $z=5.2$
\citep[][]{combes12,rawle13}, a source we selected independently from
its confirmation as a high-$z$ source through the HLS survey. We also
note that the \Planck\ selection is based on 350$\,\mu$m, 550$\,\mu$m,
and 850$\,\mu$m measurements, covering somewhat longer wavelengths
than the blind \textit{Herschel\/} surveys. The \Planck\ high-$z$
sample has a range of FIR colours, and the redshift distribution
derived from the subsample of very bright gravitational lenses does
not necessarily correspond to the redshift distribution of the overall
\Planck\ high-$z$ sample (HLS~091828.6$+$514223 at $z=5.2$ being a
case in point).

\subsection{Line profiles and luminosities}
\label{ssec:lineprofiles}

Fig.~\ref{fig:lines} shows the lines we detected in the 3-mm
band. Many sources exhibit complex line profiles, which can also be
seen in the higher frequency data (Canameras et al. 2015b, in
  prep.). These may either originate from several, gravitationally
lensed regions in the same galaxy that are blended in the large
(20\arcsec) beam of the IRAM 30-m telescope, or else they represent
sets of nearby, perhaps interacting galaxies. Alternatively, they may
represent intrinsically complex gas kinematics in single star-forming
environments, driven by the interplay of galaxy rotation, feedback, or
perhaps turbulent motion within star-forming regions. High-resolution
follow-up interferometry is currently being analysed to further
elucidate their nature.

We fitted the emission lines shown in Fig.~\ref{fig:lines} with single
Gaussian profiles to obtain the line FWHMs and integrated fluxes
listed in Table~\ref{tab:lines}. Integrated fluxes are
$I_{\rm CO}=(7{-}34)\,{\rm Jy}\,{\rm km}\,{\rm s}^{-1}$,
for ${\rm FWHM}=213$--$685\,{\rm km}\,{\rm s}^{-1}$.
A detailed analysis of
the spectral properties of our sources, using our full sets of
spectral line observations, and taking into account changes in line
ratios in individual, kinematically distinct emission line regions, will be
presented elsewhere; here we only use the spectroscopic redshifts, and
the fluxes and FWHM line widths of the lowest-$J$ CO transition observed
in each source in order to demonstrate that these are indeed gravitationally
lensed galaxies with extreme magnification factors.

In Table~\ref{tab:lines} we also compile luminosity estimates for the CO
lines that we detect.
We follow \citet{solomon97} in translating the integrated line fluxes
into CO luminosities (in brightness temperature units), setting
\begin{equation}
L^{\prime} = 3.25\times 10^7\ I_{\rm CO}
 \left(\frac{\nu}{1+z}\right)^{-2} D_{\rm L}^2 (1+z)^{-3}.
\label{eq:Lprime}
\end{equation}
Here the measured frequency $\nu$ is given in GHz, $z$ is the redshift,
$D_{\rm L}$ is the luminosity distance, the CO luminosity $L^{\prime}$
is measured in ${\rm K}\,{\rm km}\,{\rm s}^{-1}\,{\rm pc}^2$,
and the integrated CO line flux is in ${\rm Jy}\,{\rm km}\,{\rm s}^{-1}$.
In Sect.~\ref{sec:dustgasratios} we will also derive 
molecular gas mass estimates and compare with the dust
masses obtained from our SED fitting. 

\begin{table*}[htbp]
\begingroup
\newdimen\tblskip \tblskip=5pt
\caption{CO line properties obtained with EMIR on the IRAM 30-m telescope.
The columns are
source name;
CO transition;
integrated line flux;
line luminosity $\mu\ L^{\prime}$ (in brightness temperature units) derived using
Eq.~\ref{eq:Lprime}; 
molecular gas mass (see Sect.~\ref{sec:dustgasratios} for details); and 
Full width at half maximum of the emission lines.}
\label{tab:lines}
\vskip -3mm
\footnotesize
\setbox\tablebox=\vbox{
 \newdimen\digitwidth
 \setbox0=\hbox{\rm 0}
 \digitwidth=\wd0
 \catcode`*=\active
 \def*{\kern\digitwidth}
 \newdimen\signwidth
 \setbox0=\hbox{+}
 \signwidth=\wd0
 \catcode`!=\active
 \def!{\kern\signwidth}
 \newdimen\pointwidth
 \setbox0=\hbox{.}
 \pointwidth=\wd0
 \catcode`@=\active
 \def@{\kern\pointwidth}
 \newdimen\notewidth
 \setbox0=\hbox{g}
 \notewidth=\wd0
 \catcode`?=\active
 \def?{\kern\notewidth}
 \halign{\tabskip=0pt\hbox to 1.5in{#\leaderfil}\tabskip=1em&
 \hfil#\hfil\tabskip=1.5em&
 \hfil#\hfil\tabskip=1em&
 \hfil#\hfil\tabskip=0.5em&
 \hfil#\hfil\tabskip=0.0em&
 \hfil#\hfil\tabskip=1.5em&
 \hfil#\hfil\tabskip=0pt\cr
\noalign{\doubleline}
\noalign{\vskip -3pt}
\omit\hfil Source\hfil& Transition& Line flux $\mu\ I_{CO}$&
 Line luminosity $\mu\ L^{\prime}$& $\mu\ M_{\rm mol}$& FWHM\cr
\noalign{\vskip 3pt}
\omit& & [${\rm Jy}\,{\rm km}\,{\rm s}^{-1}$]&
 [$10^{11}\,{\rm K}\,{\rm km}\,{\rm s}^{-1}\,{\rm pc}^2$]& [$10^{11}\,{\rm M}_\odot$]&
 [${\rm km}\,{\rm s}^{-1}$]\cr
\noalign{\vskip 4pt\hrule\vskip 4pt}
PLCK\_G045.1$+$61.1&  4-3 & 22.9$\pm$0.3* & *6.9$\pm$0.1  & *5.6$\pm$0.1   & 213$\pm$*11\cr
PLCK\_G080.2$+$49.8&  3-2 & *9.2$\pm$0.5* & *3.2$\pm$0.2  & *2.6$\pm$0.2   & 265$\pm$*12\cr
PLCK\_G092.5$+$42.9&  4-3 & 34.3$\pm$0.2* & *9.7$\pm$0.1  & *7.8$\pm$0.2   & 453$\pm$**3\cr
PLCK\_G102.1$+$53.6&  3-2 & *5.7$\pm$1.8* & *2.4$\pm$0.7  & *2.0$\pm$0.2   & 252$\pm$*10\cr
PLCK\_G113.7$+$61.0&  3-2 & 16.5$\pm$0.2* & *5.0$\pm$0.1  & *4.0$\pm$0.2   & 528$\pm$**5\cr
PLCK\_G138.6$+$62.0&  3-2 & 22.0$\pm$1.1* & *6.8$\pm$0.3  & *5.4$\pm$0.1   & 514$\pm$*40\cr
PLCK\_G145.2$+$50.9&  3-2 & 21.9$\pm$0.8* & 12.7$\pm$0.6  & 10.2$\pm$0.2   & 685$\pm$*17\cr
PLCK\_G165.7$+$67.0&  3-2 & 25.4$\pm$0.3* & *6.8$\pm$0.1  & *5.4$\pm$0.1   & 576$\pm$**4\cr
PLCK\_G200.6$+$46.1&  3-2 & 11.2$\pm$0.1* & *4.9$\pm$0.1  & *3.8$\pm$0.1   & 458$\pm$**9\cr
PLCK\_G231.3$+$72.2&  3-2 & *9.4$\pm$0.2* & *3.9$\pm$0.1  & *3.0$\pm$0.1   & 257$\pm$**8\cr
PLCK\_G244.8$+$54.9&  3-2 & *7.4$\pm$1.0* & *3.3$\pm$0.4  & *2.6$\pm$0.2   & 382$\pm$*10\cr
\noalign{\vskip 4pt\hrule\vskip 4pt}
}}
\endPlancktablewide
\endgroup
\end{table*}

\section{Dust properties}
\label{sec:bbfit}

\begin{figure*}
\centering
\includegraphics[width=1.0\textwidth]{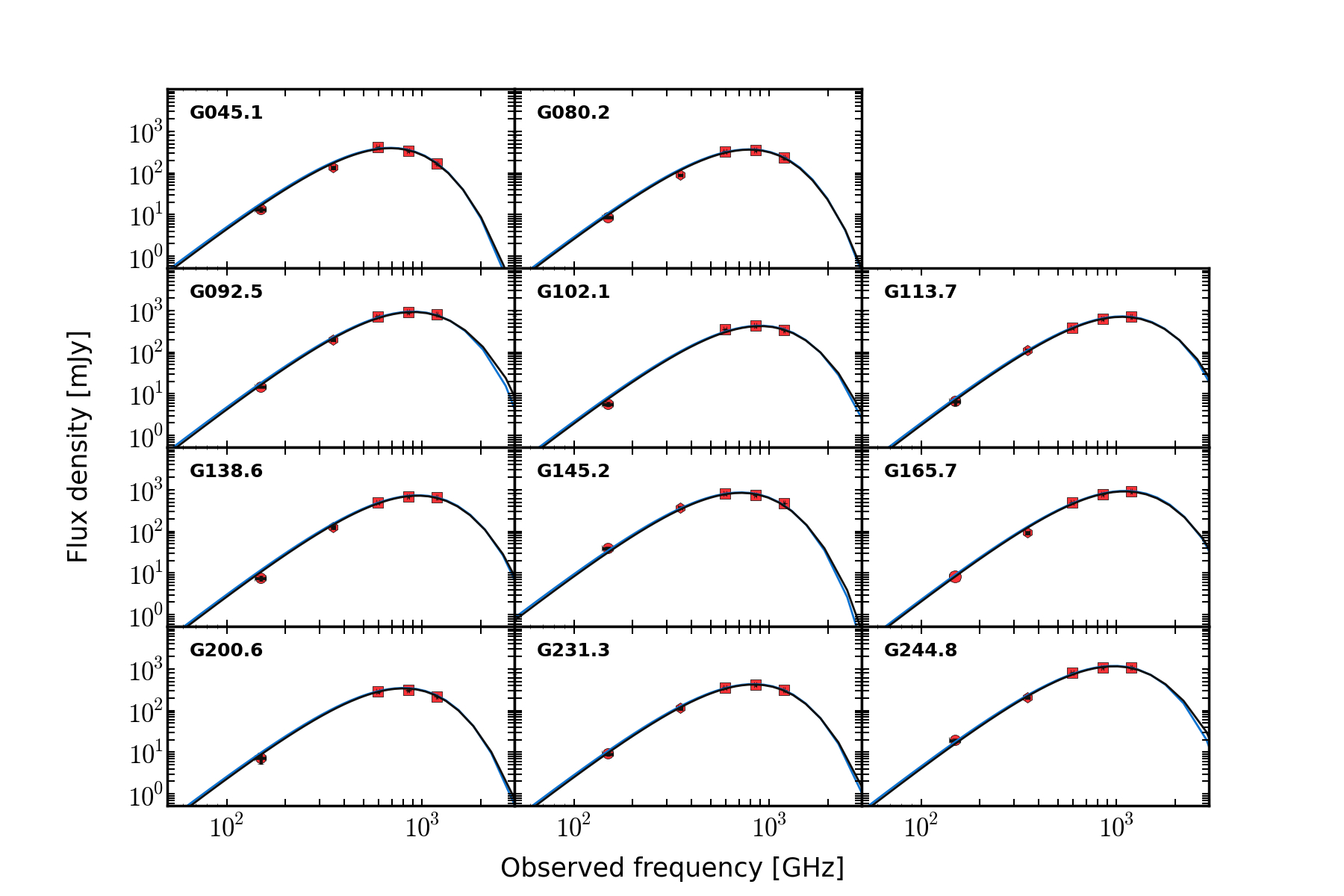}
\caption{Modified blackbody fits to the FIR-to-millimetre photometry
  of our sources as obtained with \textit{Herschel}-SPIRE at 250, 350,
  and 500$\,\mu$m, with JCMT/SCUBA-2 at 850$\,\mu$m (for eight
  sources), and with IRAM/Gismo at 2\,mm. The light and dark blue
  curves show the blackbody curves for optically thin and optically
  thick dust, respectively. Error bars are printed in black, and are
  smaller than the symbols showing the photometric data points.
\label{fig:bbfit}}
\end{figure*}

To further characterize our sources, we fitted their FIR-to-millimetre
photometry with modified blackbody distributions between 250$\,\mu$m
and 2000$\,\mu$m, using the Python {\tt curve\_fit} routine from the
scipy
package\footnote{http://docs.scipy.org/doc/scipy/reference/generated/
  scipy.optimize.curve\_fit.html}. The results are shown in
Fig.~\ref{fig:bbfit}. Most photometry points were observed with
roughly similar beam sizes between 15\arcsec\ and 30\arcsec, and we
only see single, very bright components in each image, which makes us
confident that uncertainties related to confusion and multiple sources
within the same beam do not dominate our photometry. We do see the
foreground lensing sources at short wavelengths, between the optical
and near-infrared, including the blue channels of
\textit{WISE}. However, their relatively blue colours and locations
suggest that they do not contribute significantly to the
long-wavelength emission.

In the restframe of our targets, our data cover the peak of the SED
and extend well into the Rayleigh-Jeans tail of the dust emission,
which is a particularly clean probe of the thermal emission of the
coldest dust component, dominating the overall mass budget. We
followed, e.g., \citet[][]{blain03} to allow for a frequency-dependent
optical depth of the FIR emission, which we model in the standard way
through a power law with emissivity index $\beta=$2.0 and a critical
wavelength of 100$\,\mu$m, at which the dust opacity becomes
unity. This specific choice of $\beta$ is somewhat arbitrary, but at
the signal-to-noise ratio and sparse frequency sampling of our SEDs,
the precise value has a minor impact (for plausible values of $\beta$
between 1.5 and 2.0) for our study.  For example, we find typical
offsets in dust temperature $\Delta T_{\rm d}\approx2$\,K between
$\beta=1.5$ and $\beta=2.0$, comparable to our other uncertainties.

These fits correspond to dust temperatures of $T_{\rm d}=33$--$50\,$K,
with no significant trend with redshift
(Table~\ref{tab:dust}). Uncertainties were derived from Monte Carlo
simulations, where we varied the measured flux density in each band
1000 times within the measurement uncertainties, before fitting the
SED in the same way as for the data.

The high infrared luminosities and dust temperatures of \Planck's
Dusty GEMS suggest that most of the radiation in our sources comes
from dense star-forming clouds, which is why we consider the optically
thick case more relevant to our analysis. We note, however, that had
we fit for optically thin emission, we would have derived temperatures
between $T_{\rm d}=$30\,K and 41\,K, lower by 3\,K to 11\,K in each
individual source.  Fig.~\ref{fig:bbfit} shows that all sources are
adequately fit with a single temperature component, in spite of a
slight, systematic underestimate of the 850$\,\mu$m, or alternatively,
slight overestimate of the 500-$\mu$m flux density.

\begin{figure}
\centering
\includegraphics[width=0.5\textwidth]{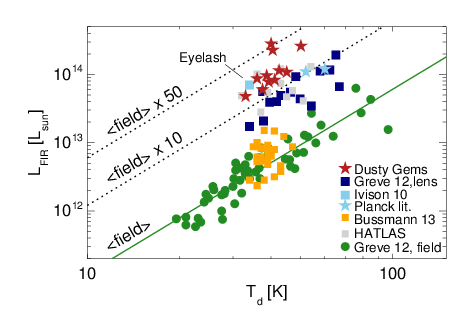}
\includegraphics[width=0.5\textwidth]{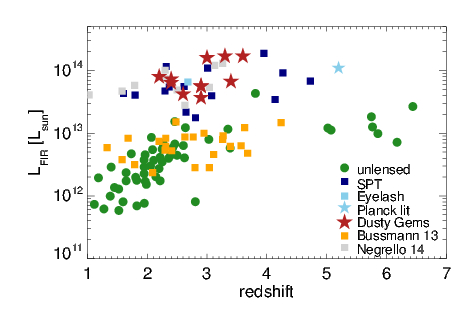}
\caption{
  Characterizing our sources by their FIR luminosities, dust temperatures,
  and redshifts.  {\it Top}: FIR luminosities versus redshifts.
  The Dusty GEMS targets are
  shown as red stars on top of the comparison samples of strongly
  lensed sources taken from the literature \citep[see][and
  references therein]{greve12}, lens candidates from the SPT survey
  \citep{vieira13},
  the Eyelash \citep{ivison10}, and the two \Planck\ lenses that were
  discovered prior to our work \citep[][]{combes12,fu12}.
  The solid line shows a fit to the general population, while the dotted lines
  show magnifications factors of 10 and 50.
  {\it Bottom}: FIR luminosity as a function of
  redshift. In the lower panel, our sources fall near the most
  luminous gravitationally lensed galaxies at similar redshifts;
  however, these are strongly dominated by AGN, like APM~08279$+$5255,
  whereas our sources are bona fide starbursts, at least in the image
  plane. This explains the greater luminosities at lower dust
  temperatures seen in the top panel.}
\label{fig:tdlfir}
\end{figure}

The temperature range we find does not differ significantly from those of
other strongly lensed samples of high-redshift dusty star-forming
galaxies selected in the FIR and (sub-)mm. This is illustrated in
Fig.~\ref{fig:tdlfir}. It is interesting that the temperature range of
our sources includes those of typical starburst galaxies at high
redshift, as well as the lower temperatures of more gradually, but
still intensely, star-forming galaxies along the main sequence
\citep[e.g.,][]{elbaz11}. We do not find an obvious trend of dust
temperature with redshift. This may either be due to the small number
of targets observed over a relatively small redshift range, or because
we are sampling environments with a range of properties within
individual galaxies.

We derive infrared luminosities by integrating over our best-fit
single-component modified blackbody SEDs between 8$\,\mu$m and
1000$\,\mu$m in the rest-frame, using the flux densities given in
Table~\ref{tab:photometry} and dust temperatures in
Table~\ref{tab:dust}. Luminosities thus obtained are in the range
$L_{\rm FIR}=(4.6{-}26.5)\times 10^{13}\mu^{-1}{\rm L}_{\odot}$, including
the unknown gravitational magnification factor $\mu$. Following
\citet{kennicutt98} we set ${\rm SFR} [{\rm M}_\odot\,{\rm yr}^{-1}]$
$=4.5\times 10^{-37}\,L_{\rm FIR}$ [W], finding star-formation rates
${\rm SFR}=8000$--$46000\,\mu^{-1}\,{\rm M}_{\odot}\,{\rm yr}^{-1}$,
suggesting high intrinsic star-formation rates and high magnification
factors. Corresponding apparent dust masses are $M_{\rm
  d}=(3.5{-}11)\times 10^{9}\, \mu^{-1}{\rm M}_{\odot}$
(Table~\ref{tab:dust}).

Fig.~\ref{fig:bbfit} shows that we cover the peak of the dust SED for
most sources, but nonetheless our sensitivity to warmer dust
components probed in the FIR shortwards of 250$\,\mu$m (60--80$\,\mu$m in
the rest-frame) is limited, and we are likely to miss emission from
warmer dust components. Through template fitting with {\tt DecompIR} 
\citep[][]{mullaney11}, which we will further discuss in
Sect.~\ref{sec:templatefit}, we derived fiducial correction factors
between the modified blackbody fits obtained here, and the
templates we used to accommodate our FIR-to-submillimetre photometry
(and constraints from \textit{WISE\/} at 22$\,\mu$m within the same
SED fit). We thus find that our modified blackbody fits may
underestimate the IR luminosity of our targets by 10--30\,\%. We also
list the respective values obtained from {\tt DecompIR} in
Table~\ref{tab:dust}. Taken together, the direct estimate from the
modified blackbody fits, and those from the templates, set a
plausible range of luminosities. These luminosities would need to
be corrected towards larger values if an additional warmer dust
component was found with infrared constraints shortward of
250~$\mu$m.

We follow, e.g., \citet{greve12} in estimating dust masses, $M_d$,  by setting 

\begin{equation}
M_d = \mu^{-1} \frac{D_L^2 S_{\nu_0}}{(1+z)\kappa_{\nu_r}} \left( B_{\nu_r}(T_d)-B_{\nu_r}(T_{CMB}(z))\right)^{-1}
\end{equation}

where $D_L$ is the luminosity distance to redshift $z$, $S_{\nu_0}$
the flux density at the observed frequency $\nu_r(1+z)^{-1}$, which we
set to 600~GHz, (corresponding to the 500~$\mu$m band of {\tt
  SPIRE}). $T_{CMB}$ is the CMB temperature at redshift $z$. We used
$\kappa_{\nu_r}/ \left(m^2\ kg^{-1}\right)\ =\ 0.045\ \times\ (\nu_r/\ 250\ GHz)^{\beta}$,
with $\beta =2.0$ as already stated above. With this approach and
these assumptions, we find dust masses of $M_d = (3.5-11)\times 10^9
\mu^{-1} $M$_{\odot}$, with the unknown gravitational magnification
factor $\mu$. Results for individual sources are listed in
Table~\ref{tab:dust}.

\section{Signatures of gravitational lensing}
\label{sec:lensing}

\subsection{Flux densities and morphologies}
The first observational hint that our targets might indeed be strongly
gravitationally lensed galaxies is their sheer brightness and small
angular size below 18\arcsec (the FWHM of the SPIRE point spread
function at the highest frequency, 250$\,\mu$m), whereas 90\% of
 the 234 Planck high-z candidates with Herschel/SPIRE follow-up
 consist of multiple sources. Unlensed high-redshift galaxies with
 FIR flux densities as high as those of our sources are very rare.
 For example, for flux densities $S_{350} >$ 400 mJy at 350$\mu$m and
 near the peak of the FIR SEDs of most GEMS, the \citet{bethermin12} models 
 predict 3 sources sr$^{-1}$ at z$\ge$2 that are not, and 53
 sources sr$^{-1}$ that are gravitationally lensed. For a given source
 at these flux densities, it is therefore much more likely to find a
 gravitationally lensed galaxy than a galaxy in the field. We
 emphasize that this is true only for galaxies with confirmed high
 redshifts like the GEMS. The overall population of bright FIR
 galaxies is at much lower redshifts and has bluer FIR colors.

Given that more typical high-z galaxies in the field have FIR
  flux densities of few 10s\,mJy, the observed flux densities of our
  sources make us expect magnification factors of at least a few and
perhaps greater than 10, even if these were intrinsically amongst the
most luminous sources on the sky. For more moderate, more typical
intrinsic flux densities, the gravitational magnification would be
accordingly higher. The one alternative interpretation could be that
these are multiple, highly concentrated galaxies within projected
distances of a few tens of kpc, so that they are not resolved into
individual galaxies by the SPIRE beam, which corresponds to about
120\,kpc at $z\approx2$ \citep[for an example see][]{ivison13}.

More direct evidence that our sources indeed have a lensing nature
comes from their morphologies. Strongly gravitationally lensed
galaxies may either be single or multiple compact images seen near the
intervening foreground lensing source, giant arcs extending over
several arcseconds, or even partial or complete Einstein
rings. However, reaching the required spatial resolutions of about
1--2\arcsec\ or better requires either deep optical or near-infrared
imaging for sources that are not too heavily obscured, or,
alternatively, submm or millimetre interferometry.

Fig.~\ref{fig:morphologies} shows the \textit{Herschel\/} image as
contours overlaid on optical imaging, and Fig.~\ref{fig:morphsma}
the dust morphology  obtained with the SMA at
850$\,\mu$m. These figures clearly show
that our sources are either single, compact objects at
2\arcsec\ resolution (as is the case for, e.g., PLCK\_G244.8$+$54.9
and PLCK\_G138.6$+$62.0), or giant arcs (e.g., PLCK\_G145.2$+$50.9 and
PLCK\_G080.2$+$49.8). PLCK\_G244.8$+$54.9 and PLCK\_G138.6$+$62.0
have flux densities of 1054$\pm$10 and 664$\pm$8~mJy, respectively,
making it implausible that these are individual high-redshift
galaxies if not benefitting from a strong boost from gravitational
lensing through a massive foreground source. At a 2\arcsec\ beam,
intrinsic source sizes of unlensed galaxies would be at most 16~kpc,
making it implausible to see a small group of high-redshift galaxies
within a single beam.

Our sources are associated with overdensities of massive intervening
galaxies at intermediate redshift, either galaxy groups or
clusters. Table~\ref{tab:photometry} lists the redshifts of the
brightest intervening galaxies along the line of sight to our
targets, which were taken from the SDSS. They are either photometric
redshifts or (in a few individual cases highlighted in
Table~\ref{tab:photometry}), spectroscopic redshifts. A detailed
analysis of the lensing structure based on our own proprietary
optical/NIR photometry obtained at the CFHT and the VLT is on-going.
The only target associated with a foreground object for which we
currently have no good redshift estimate is G145.2$+$50.9, which forms
a near-complete Einstein ring with a diameter of 10\arcsec\ around a
very red foreground source, which we currently only detect in the
J-band.  For this galaxy, our optical/NIR photometry is not yet
complete, and we must therefore defer a detailed analysis of the
foreground source to a later publication.

In the rest of this section we will present additional empirical
evidence to illustrate that all of our sample consists of strongly
gravitationally lensed galaxies, including those for which we do not
yet have morphological constraints.

\subsection{FIR dust continuum} 
\label{ssec:dust}
\citet{greve12} suggested the use of the relationship between $T_{\rm
  d}$ and $L_{\rm FIR}$ as an empirical indicator of whether a source
is strongly gravitationally lensed. According to the Stefan-Boltzmann
law, the luminosity emitted by a modified blackbody at a given
temperature depends only on the size of the emitting surface and the
emissivity.  Therefore, if our galaxies form a subset of the generic
population of dusty, intensely star-forming high-redshift galaxies,
the gravitational magnification factor should lead to an apparent
increase in dust luminosity at a given temperature for gravitationally
lensed galaxies, compared to unlensed galaxies.

In Fig.~\ref{fig:tdlfir} we show where our sources fall on a plot of
dust temperature versus FIR luminosity, compared to the galaxies of
\citet{greve12} and other sources from the literature. It is
immediately clear that our sources are significantly brighter at a
given temperature than galaxies in the field, and even brighter than
most of the previously known gravitationally lensed galaxies. The
dashed lines illustrate luminosities that are 10 times and 50 times
brighter than expected from a simple least square fit
to the population of field galaxies (green solid line). We highlight
the Cosmic Eyelash of \citet{swinbank10} and the two previously
confirmed gravitationally lensed galaxies with \Planck\
counterparts \citep{fu12,combes12}.

For a more explicit, although not necessarily very precise estimate
that exploits this relationship between dust temperature, luminosity,
and source size, we can use the Stefan-Boltzmann law directly to infer
the magnification factor from the observed luminosity and dust
temperature. The approach is described in detail in
\citet{greve12}. Given the somewhat higher dust temperatures of our
lens candidates compared to the SPT sources of \citet{greve12}, we
have a larger range of ``effective'' Stefan-Boltzmann constants (which
correct for the lower emissivity of a modified relative to a genuine
blackbody), $\sigma_{\rm eff}/\sigma=0.5$--0.65, and apparent emitting
surfaces that are between 8 and 40 times larger than those of typical FIR
galaxies, implying magnification factors that are roughly
similar. These results are also shown in Fig.~\ref{fig:tdlfir}. We note
that the estimates are very uncertain, since they assume
that each high-redshift galaxy has a similar emissivity, that the dust
becomes optically thick at similar wavelengths, and that the
interstellar medium (ISM) of
the galaxies is dominated by a uniform dust component with a single
temperature. Each of these assumptions can only be approximately true,
and this will lead to uncertainties of at least factors of a few in
this estimate. This is illustrated by the magnification factor of
about 10 of the ``Cosmic Eyelash,'' which we constrained with the same
approach, and which falls into a very similar region of the diagram as
our sources. The ``Cosmic Eyelash'' has intrinsic magnification
factors of 20$-$60 \citep[][]{swinbank11}, with an average factor of
32.5. In spite of these caveats, Fig.~\ref{fig:tdlfir} provides 
 additional indirect support that our sources are indeed strongly
gravitationally lensed galaxies.

\subsection{Molecular gas lines}
\label{ssec:mol}

\begin{figure}
\centering
\includegraphics[width=0.5\textwidth]{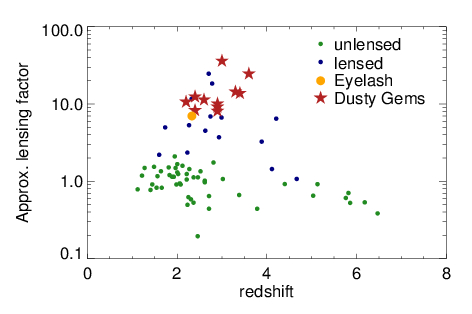}
\includegraphics[width=0.5\textwidth]{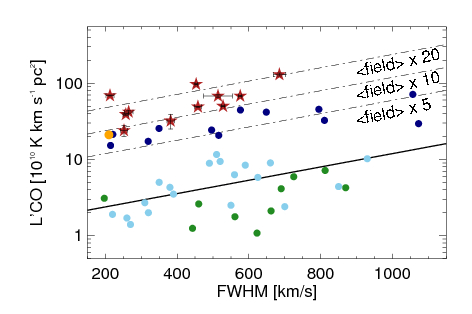}
\caption{{\it Top}: Order-of-magnitude estimate of the gravitational
  magnification factor derived from the dust temperature and FIR
  luminosity of our sources compared to FIR/submm galaxies in the
  field (see Sect.~\ref{ssec:dust} for details). {\it Bottom}: CO
  emission-line luminosity as a function of FWHM line width, using the
  same symbols as in the top panel, with additional light blue
    dots showing those sources of \citet{bothwell13}, which are neither
    gravitationally lensed nor AGN hosts. 
The solid and dashed lines show
  the best linear fit to the sample of field galaxies, as well as
  shifts of this line corresponding to magnification factors of 5, 10,
  and 20. Error bars show the 1$\sigma$ uncertainties of our
    single-component Gaussian line fits.}
\label{fig:colens}
\end{figure}

More empirical evidence for strong gravitational magnification of our
galaxies comes from the emission-line properties that we measured with EMIR
on the IRAM 30-m telescope. To demonstrate that their galaxies
discovered through the H-ATLAS survey are indeed strongly
gravitationally lensed, \citet{harris12} compiled the
CO(1-0) line luminosities $L^{\prime}$
and FWHMs of high-redshift galaxies, finding that typical
galaxies follow a broad trend between luminosity and line width
(Fig.~\ref{fig:colens}). Their own galaxies and other strongly
gravitationally lensed galaxies from the literature stand out by 1--2
orders of magnitude above this relationship, owing to the boost in
line luminosity by the gravitational lens. In contrast, lines are
expected to be as broad as in field galaxies (or narrower, 
because smaller regions of the large-scale velocity field of the
galaxy are being observed). If a small region in a galaxy is magnified by
strong gravitational lensing, then only parts of the rotation curve
will be sampled, and turbulent motion will also be smaller. Therefore,
observed line widths should be narrower than if galaxy-wide
radii are probed, as in field galaxies.

Fig.~\ref{fig:colens}, inspired by Figure~7 of \citet{harris12}, shows
where our sources fall relative to the Harris et al.\ sample and those
in the literature. 
Our sources span a large range in FWHM in this diagram,
but are all clearly within the regime of gravitationally lensed
galaxies. The FWHMs of some of our targets may be as large as those of
overdensities of multiple lensed sources that are at best moderately
gravitationally lensed
like the example in \citet{ivison13}. Their luminosities are
factors of a few brighter, placing them firmly above the
field galaxies, and into the lens regime. The example of the Cosmic
Eyelash, which has a magnification factor of 20--60 shows however that
the lensing factor derived from the position of a source in this diagram can
only be an approximate indicator. The line luminosities and
  profiles provide additional evidence in support for the lensing
  hypothesis.

\section{Dust heating and AGN content}
\label{sec:templatefit}

Many dusty high-$z$ starburst galaxies host powerful AGN.  For example
\citet{alexander05} find that 75\,\% of submillimetre galaxies have AGN
detected at X-ray wavelengths \citep[but see also][ who find
  significantly lower fractions of 10--20\,\%]{laird10}. Powerful,
obscured AGN could contribute significantly to $L_{\rm FIR}$, which
would lower the magnification factors that are needed to explain the
extraordinary apparent brightness of our targets. The inferred
dust temperatures of our sources are akin to those of other high-$z$
starburst galaxies and would be very low compared to those of
obscured quasars, which are typically above 70\,K, often far above.
However, our estimates may be biased towards low temperatures
because of the absence of mid-infrared constraints shortward of
50-80$\mu$m in our data sets. This makes it worthwhile to
investigate in more depth what the possible AGN contribution might
be to the infrared luminosity budget of our targets.

Not all AGN play a significant role for the total infrared luminosity
budget of their host galaxies. Ideally, searching for obscured AGN
activity in our targets would require PACS or at least deep {\textit
  Spitzer\/} 24-$\mu$m photometry, to sample the dust SED shortward of
about 100$\,\mu$m in the rest-frame. Since our sample was only
discovered with SPIRE in the last weeks of the \textit{Herschel\/}
mission, such data sets are not available.  Additionally, in spite of
their exceptional brightness (for high-redshift targets), they are
still too faint to obtain robust mid-IR photometric constraints with
\textit{SOFIA}. We might therefore not be able to give tight
  constraints on the presence of weak AGN activity in all of our
  sources, however, most important for our purposes here is to show
  that AGN radiation is not energetically dominant in our sources
  (i.e., they do not contribute $> 0.5\ L_{bol}$), which would have
  important implications for the estimated star formation rates,
  plausible ranges of intrinsic infrared luminosities, and gas
  excitation.

\subsection{WISE 22$\mu$m photometry}

The \textit{Wide-Field Infrared Survey Explorer} (\textit{WISE\/}) in the
22-$\mu$m band covers rest-frame wavelengths between 4 and 7$\,\mu$m
for our sources, i.e., the far blue tail of hot dust emission from
powerful AGN. Consequently, \textit{WISE\/} has been successfully used
to identify heavily obscured AGN at high-$z$ \citep[e.g.,][]{yan13}. All
of our sources are bright enough to be detected in \textit{WISE\/}
imaging at 22$\,\mu$m, and seven are included in the
  \textit{WISE\/} catalogues. The remaining four can be seen as very
  faint sources in the \textit{WISE\/} images, but are not bright
  enough to obtain robust flux measurements. We will in the following
  use the flux measurements from the \textit{WISE\/} catalogs and
  treat the 5$\sigma$ rms of this catalog at 5.4~mJy as upper limit
  for the fainter sources.  Individual flux densities are listed for
  all sources in Table~\ref{tab:photometry}.

Although we do detect nearby bluer sources in each case at
  3.4$\,\mu$m, 4.6$\,\mu$m, and 12$\,\mu$m, the spectral energy
  distribution in all but one source falls off too steeply in the
  three blue \textit{WISE\/} bands to contribute significantly to the
  22$\,\mu$m detections. Their colours suggest that these are
  low-to-intermediate-redshift objects, which we consider part of the
  intervening lensing structure.

\subsection{Constraining AGN dust heating with {\tt DecompIR}}
We used the publicly available package {\tt DecompIR}
\citep{mullaney11} to constrain the potential contribution from AGN
emission to the infrared SED, using our six available data points
between 22$\,\mu$m and 2\,mm, as listed in
Table~\ref{tab:photometry}. {\tt DecompIR} performs $\chi^2$ fits to
infrared SEDs using composites of empirically constructed templates of
starburst galaxies and AGN. We only have six photometric data points,
but we sample the Rayleigh-Jeans tail and the peak of the spectral
energy distribution of the emission from the coldest dust component,
and therefore constrain the most abundant dust component
well. We have spectroscopic redshifts and hence can estimate robust dust
temperatures, and therefore we consider this part of the SED
to be well constrained. Our main uncertainties are blueward of the
dust peak, which we only sample with the 22$\,\mu$m data point from
\textit{WISE} and the 250~$\mu$m data point from SPIRE. The
250~$\mu$m band corresponds to 50-80~$\mu$m in the rest-frame for our
targets, is slightly blueward of the peak of the dust emission
in most of our sources (Fig.~\ref{fig:bbfit}), and falls near the
expected peak of the infrared emission from hot dust in the AGN torus
in most templates \citep[e.g.,][]{polletta07, nenkova08}. The SEDs of
powerful AGN host galaxies at high redshifts are typically dominated
by AGN emission at these wavelengths
\citep[e.g.,][]{sajina12,drouart14}.

The {\tt DecompIR} software package provides a small set of SED templates,
derived from nearby starburst galaxies. One complication in using these
templates for our sources is that we find a range of dust temperatures, which
are not all well represented by existing templates. This would lead to
considerable (but somewhat artificial) discrepancies in the template
fitting that would be hard to overcome without a self-consistent model of
the mid-to-far-infrared SED of high-redshift galaxies, which does not
yet exist. To avoid such mismatches, while make optimal use of our
existing constraints without overinterpreting the mid-IR data,
we therefore constructed a simple starburst template for each of our
galaxies from the ``SB2'' template of {\tt DecompIR}
\citep[corresponding to the spectral energy distribution of
  NGC7252,][]{mullaney11}, but correcting for the mismatch in dust
temperatures.  We therefore fitted and removed from the template the
modified blackbody contribution from cold dust in the FIR. The
residual is a template of the SED in the mid-infrared only, to which we
added the modified blackbody emission obtained from our FIR fits,
using the measured temperature for each source. We selected this
particular template, because it provided the best match to the FIR
and sub-millimeter measurements of the dust peak and Rayleigh-Jeans
tail, with no regard of the goodness of fit of the 22~$\mu$m
observations.  For the AGN component, we simply used the
\citet{polletta07} type-2 QSO template which is already part of {\tt
 DecompIR}.

\begin{figure*}
\centering
\includegraphics[width=1.1\textwidth]{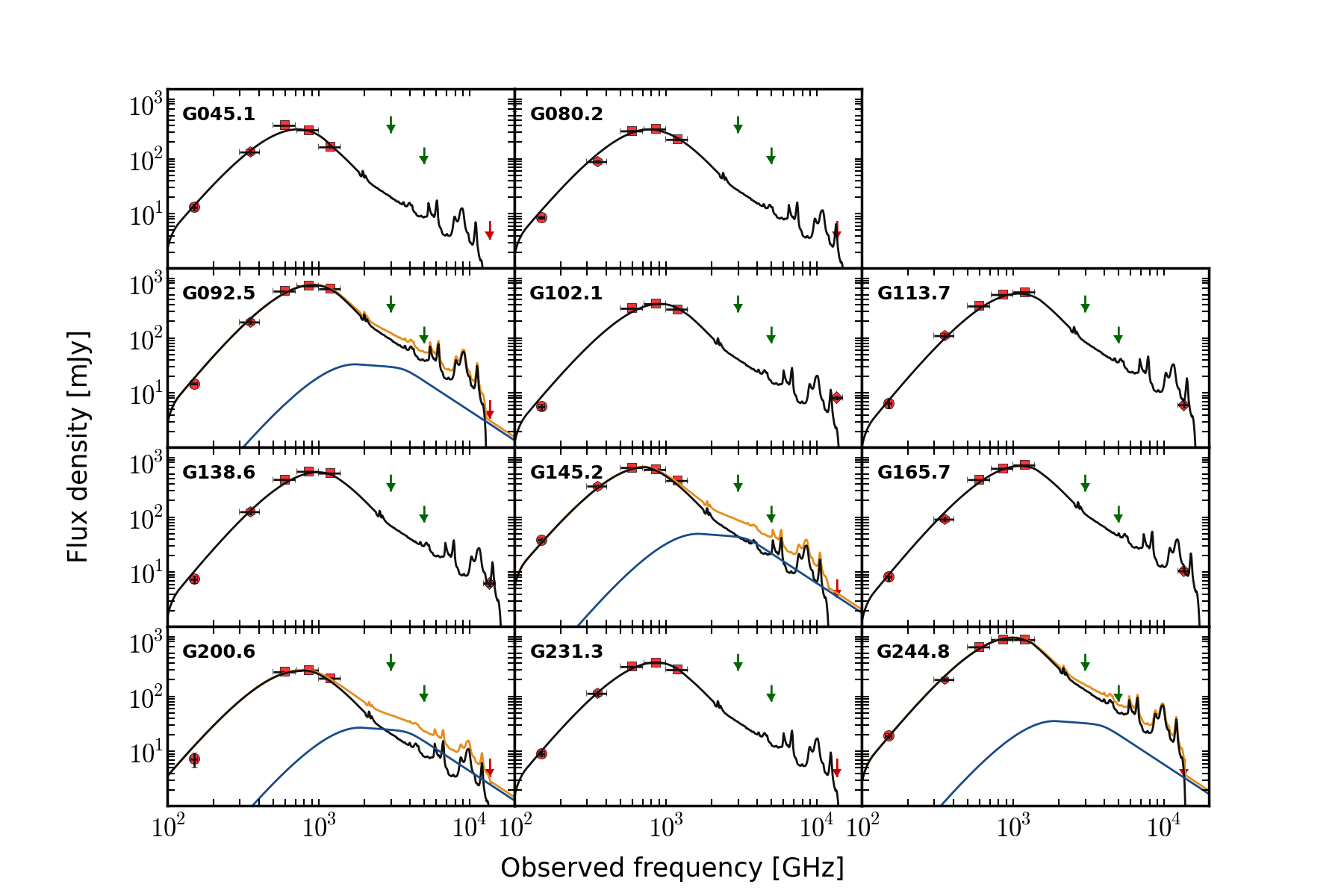}
\caption{Infrared-to-millimetre spectral energy distributions of all
  of our 11 sources obtained using {\tt DecompIR} \citep{mullaney11}. 
  Red dots show our data points, where the error bars along the
    abscissa indicate the width of each band. The error bars along the
    ordinate are often smaller than the symbol size. Green downward
    arrows show the 90\% completeness limit of the IRAS all-sky survey
    at 60 and 100 $\mu$m, respectively. Red downward arrows at
    22$\mu$m show the 5$\sigma$ upper limits provided by the WISE
    catalogs, where the counterparts were fainter than the 5.4~mJy flux limit of the WISE catalog. Black, blue, and yellow
    lines show the starburst and AGN component (if an AGN was
    fitted), and the sum of both, respectively.}
\label{fig:decompir}
\end{figure*}

Fig.~\ref{fig:decompir} shows that we can rule out a bolometrically
dominant AGN contribution from buried quasars, in all cases. Only
three galaxies have best-fit results formally inconsistent with pure
starbursts. In PLCK\_G092.5$+$42.9 and PLCK\_G244.8$+$54.9, the
putative AGN contribution to the FIR luminosity is below 10\,\%, and
it is below 30\,\% in PLCK\_G145.2$+$50.9. In all other cases, {\tt
  DecompIR} finds the best fit for a pure starburst
SED. 

\begin{figure}
\centering
\includegraphics[width=0.48\textwidth]{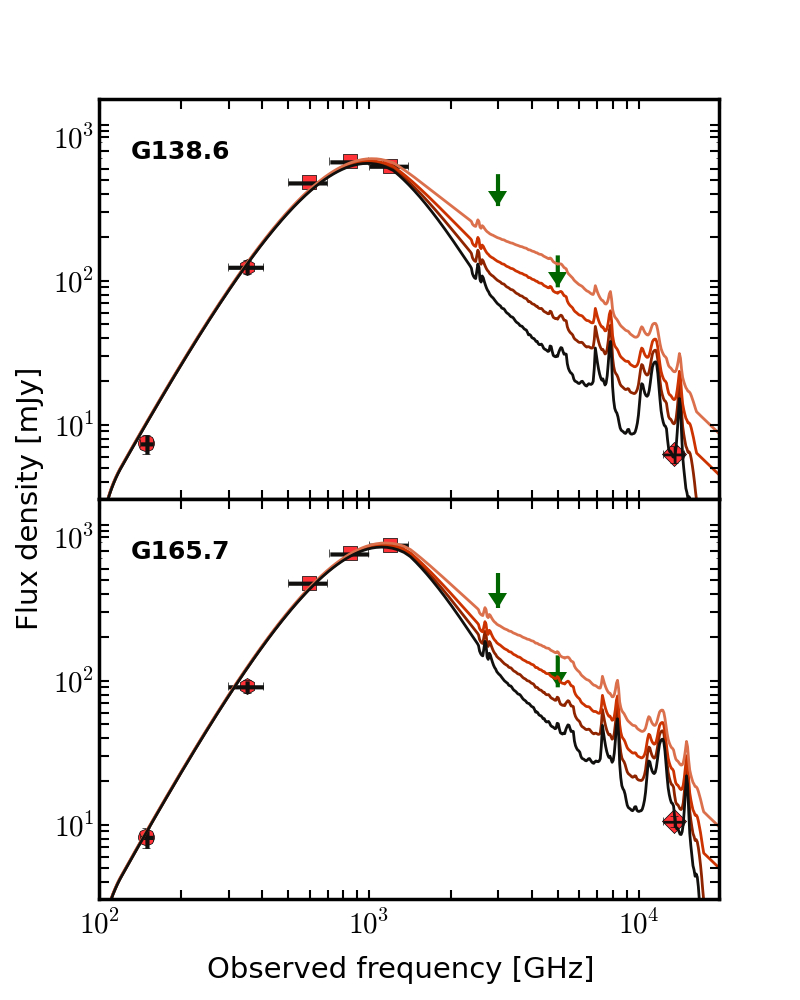}
\caption{A schematic illustrating the the impact of 0.1, 0.3, and
 $0.5\times\ L_{FIR}$ of AGN contamination to the shape of the SED of
 our sources, on the example of G138.6$+$62.0 and G165.7$+$67.0. 
}
\label{fig:noagn}
\end{figure}

Fig.~\ref{fig:decompir} also shows the 90\% completeness limits
of the IRAS faint source survey \citep[][]{moshir92}, 120~mJy and
440~mJy at 60~$\mu$m and 100~$\mu$m, respectively (green downward arrows). Although these limits
are not constraining for the SED fits that we obtained with SPIRE,
SCUBA-2, GISMO, and WISE (and were therefore not included in our
  {\tt DecompIR} fits) they do illustrate that a hypothetical bright
quasar component would have led to a 60~$\mu$m detection at least in
our brightest targets. In Fig.~\ref{fig:noagn} we illustrate the
impact of fiducial AGN contribution of 0.1, 0.3, and $0.5\times
L_{FIR}$ on the example of G138.6$+$62.0, a source with rather average
FIR brightness and 250-to-350~$\mu$m color in our sample.

Obviously, these fits are uncertain, given that they cover the far
blue wing of the dust SED expected from bright AGN, and a spectral
range that has important contributions from rich mid-infrared spectral
features, in particular from Polycyclic Aromatic Hydrocarbons (PAH),
which dominate the infrared spectral energy distributions of starburst
galaxies between about 3$\mu$m and 10$\mu$m \citep[e.g.,][]{armus07}.

The equivalent widths of the PAH bands have a wide dispersion in
high-redshift galaxies \citep[e.g.,][]{fiolet10}, and are
generally larger than those measured in low-redshift ULIRGs,
including NGC~7252, which we used as template in our {\tt DecompIR}
fits. Additional uncertainties come from the stellar continuum,
which reaches comparable strength to the dust continuum in the
wavelength range covered by the 22~$\mu$m observations, about
5-7~$\mu$m, and whose shape depends strongly on the specific
star-formation rate, star-formation history, and geometry of dust
obscuration in the host galaxy. Compared to the average mid-infrared
SED of starburst galaxies \citep[][]{brandl06}, the stellar continuum
in NGC~7252 seems relatively faint, and \citet{sajina12} also found
that the mid-IR continuum of high-redshift starburst galaxies seems to
be brighter than in low-redshift galaxies.  Other spectral features
might also contribute to boosting the observed 22~$\mu$m flux density.
For example \citet{fiolet10} report the detection of warm H$_2$ line
emission at 6.9~$\mu$m in the rest-frame. Moreover, the
 foreground galaxies may also contribute to the 22~$\mu$m flux, given the
 12\arcsec\ beam of WISE. However, all these uncertainties would
only act to lower the putative intrinsic contribution of AGN heating
relative to what we observe. It will of course be very interesting to
study the mid-infrared spectral properties of our sources in depth
once that JWST/MIRI will become available.

Relying upon these arguments, we conclude that the
dust heating in our sources is not dominated by heavily obscured
quasars, but that the \Planck\ Dusty GEMS are genuine starburst
galaxies with at most a minor contribution of AGN heating to the dust
emission and the bolometric energy budget of our sources. Of course
this only refers to the regions that we are seeing under the
gravitational microscope of the foreground lensing systems. The global
energy budgets of AGN and star formation in these galaxies may be
different, if the AGN lies along a line of sight that is not strongly
magnified \citep[see also][]{serjeant12}.

\section{FIR-radio correlation}
\label{sec:FIRradiocorrel}

Given the extraordinary brightness of our targets on the sky, we can
use the 1.4\,GHz VLA survey of the northern sky, FIRST
\citep[][]{becker95}, to search for counterparts of our sources at an
observed frequency of 1.4\,GHz (between 4.5 and 6.4\,GHz in the
rest-frame). We find that six of our sources have counterparts within
5\arcsec\ of a source in the FIRST catalogue, and the other five have
sensitive upper
limits (see Table~\ref{tab:first}). Here 5\arcsec\ corresponds to the
beam size of FIRST. We assume a radio spectral index $\alpha_{\rm 1.4}=-0.8$
\citep[as is appropriate for star-forming regions,][]{condon92}
to convert these flux densities to a monochromatic
radio flux density at rest-frame 1.4\,GHz, as well as to estimate a
radio power.  For galaxies without detections we use the 3$\,\sigma$
upper limits implied by the rms given in the FIRST catalogue. Results
for individual sources are listed in Table~\ref{tab:first}. All
  sources are listed in the FIRST catalog as marginally resolved,
with deconvolved sizes between 1.5\arcsec\ and 7\arcsec. The
  5\arcsec\ beam of FIRST is not always sufficiently small to rule out
  contamination from a radio nucleus in the foreground lensing
  galaxy. However, most sources are isolated enough to conclude that
the radio counterpart is at the position of the high-redshift galaxy,
and not coincident with a massive, intermediate-redshift galaxy that
could host a radio-loud AGN. This is particularly the case for
PLCK\_G244.8$+$54.9.

\begin{table}[htbp]
\begingroup
\newdimen\tblskip \tblskip=5pt
\caption{Radio continuum properties extracted from the FIRST catalogue.
The columns are: source name; relative distance between the SPIRE
position at 250$\,\mu$m and the position of the FIRST counterpart, if
detected within 5\arcsec; integrated flux density at 1.4\,GHz from the
FIRST catalogue and gravitational magnification factor of the radio
component, $\mu_{1.4}$; radio luminosity at 1.4\,GHz in the rest-frame
and gravitational magnification factor of that component, $\mu_{1.4}$,
for sources without 1.4 GHz detections, the luminosities are derived
from 3$\times$ the rms given in column 3; the Ratio between radio and
FIR luminosity (see Sect.~\ref{sec:FIRradiocorrel}), i.e.\ the
$q$-parameter.}
\label{tab:first}
\vskip -5mm
\footnotesize
\setbox\tablebox=\vbox{
 \newdimen\digitwidth
 \setbox0=\hbox{\rm 0}
 \digitwidth=\wd0
 \catcode`*=\active
 \def*{\kern\digitwidth}
 \newdimen\signwidth
 \setbox0=\hbox{+}
 \signwidth=\wd0
 \catcode`!=\active
 \def!{\kern\signwidth}
 \newdimen\pointwidth
 \setbox0=\hbox{.}
 \pointwidth=\wd0
 \catcode`@=\active
 \def@{\kern\pointwidth}
\halign{\tabskip=0pt\hbox to 1.4in{#\leaderfil}\tabskip=0.5em&
 \hfil#\hfil\tabskip=0.5em&
 \hfil#\hfil\tabskip=0.0em&
 \hfil#\hfil\tabskip=0.5em&
 \hfil#\hfil\tabskip=0pt\cr
\noalign{\doubleline}
\omit\hfil Source\hfil& Dist.& $\mu_{1.4}\ S_{1.4}^{\rm int}$& $\mu_{1.4}\ L_{\rm 1.4}$& $q$\cr
\noalign{\vskip 3pt}
\omit& [arcsec]& [mJy]& [$10^{25}\,{\rm W}\,{\rm Hz}^{-1}$]& \cr
\noalign{\vskip 4pt\hrule\vskip 4pt}
PLCK\_G045.1$+$61.1&    &       $<$0.135& *$<$3.2& $>$2.4\cr
PLCK\_G080.2$+$49.8&    &       $<$0.135& *$<$1.8& $>$2.4\cr
PLCK\_G092.5$+$42.9& 0.9&  1.50$\pm$0.16& 11.2& 2.35$\pm$0.11\cr
PLCK\_G102.1$+$53.6&    &       $<$0.149& *$<$2.5& $>$2.5\cr
PLCK\_G113.7$+$61.0& 3.0&   1.9$\pm$0.14& *6.9& 2.2$\pm$0.08\cr
PLCK\_G138.6$+$62.0& 2.7&  2.01$\pm$0.16& *7.3& 2.1$\pm$0.08\cr
PLCK\_G145.2$+$50.9&    &       $<$0.144& *$<$3.9& $>$2.8\cr
PLCK\_G165.7$+$67.0& 4.1&  3.41$\pm$0.15& 10.1&  2.0$\pm$0.04\cr
PLCK\_G200.6$+$46.1& 1.1&  1.23$\pm$0.14& *6.9& 1.9$\pm$0.11\cr
PLCK\_G231.3$+$72.2&    &       $<$0.151& *$<$2.5&  $>$2.5\cr
PLCK\_G244.8$+$54.9& 3.2& 2.26$\pm$0.14 & 13.6& 2.3$\pm$0.06\cr
\noalign{\vskip 4pt\hrule\vskip 4pt}
}}
\endPlancktablewide
\endgroup
\end{table}

The far-infrared radio correlation is commonly parametrized as
$q=\log{L_{\rm IR}/(3.75\times10^{12}\,{\rm W})}
-\log{ L_{\rm 1.4}/({\rm W}\,{\rm Hz}^{-1})}$,
where $L_{\rm IR}$ is measured in the range
8--1000$\,\mu$m in the rest-frame, and the centimetre continuum at a
rest-frame frequency of 1.4\,GHz. Early \textit{Herschel\/} results
suggest $\langle q \rangle =2.4\pm0.12$ in the local Universe
\citep[][]{jarvis10}, although values can change considerably even
within individual galaxies \citep[e.g.,][]{tabatabaei13}. Towards
higher redshifts, results show less of a consensus.  \citet{kovacs06},
\citet{vlahakis07}, \citet{michalowski10}, and \citet{bourne11} report
an increase in radio power at a given FIR luminosity at $z\ga1$, with
$\langle q \rangle =2.0$. However, several studies
\citep[e.g.,]{ivisonBLAST,ivison10b,sargent10,thomson14}
find average values in the range $q=2.4$--2.7,
out to $z\approx2$. Most of these studies use stacking analyses
to overcome the inherent faintness of high-redshift galaxies in the
FIR/submm and in the radio, and are based on ground-based data that
do not sample the FIR dust SED very well. In the Cosmic Eyelash, a
single, strongly lensed galaxy at $z=2.4$ with \textit{Herschel}-SPIRE
coverage, \citet{ivison10} find $q=2.4$.

\begin{figure}
\centering
\includegraphics[width=0.50\textwidth]{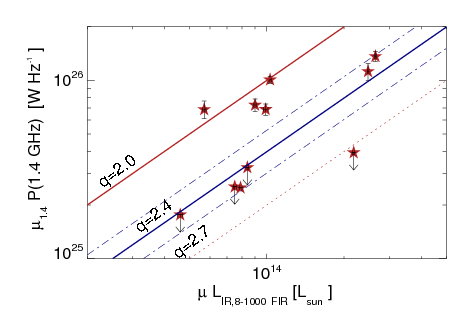}
\caption{Far-infrared radio correlation of our sources, as discussed in
  Sect.~\ref{sec:FIRradiocorrel}.  $\mu_{FIR}$ and $\mu_{1.4}$
    correspond to the magnification factors appropriate for the dusty
    and 1.4~GHz synchrotron emission, respectively. Solid blue and red
  lines show the relationship for $q=2.4$ and 2.0, respectively, while
  the red dotted line if for $q=2.7$. The dot-dashed lines show the
  $\pm1\,\sigma$ scatter about $q=2.4$ at low redshift from
  \citet{jarvis10}.
\label{fig:firradiocorrelation}}
\end{figure}

In Fig.~\ref{fig:firradiocorrelation} we show where our sources fall
relative to the local FIR-radio correlation,  assuming that
differential lensing between the FIR and radio synchrotron emission
does not play a major role in our case. Four of our sources are
brighter by up to about 0.4 dex in the radio, and fall outside the
1$\sigma$ scatter around the distribution of $q$ parameters of
low-redshift galaxies. Six galaxies fall within 1$\sigma$ from this
relationship, however, three of these galaxies only have upper
limits on $q$, and would fall outside the 1$\sigma$ scatter if we
had adopted 1$\sigma$, instead of 3$\sigma$ upper limits. One galaxy has
q$>$2.7 at 3$\sigma$.

If
the 1.4\,GHz emission is caused by star formation, then the lower $q$
values cannot be caused by our assumption of the spectral index
$\alpha_{1.4}=-0.8$. Only for flat radio spectral energy distributions
($\alpha_{1.4}=-0.2$ or above, which would only be possible for an
AGN) would our highest-$q$ sources be consistent with the
FIR-correlation and $q=2.4$.  If this were the case, our sources with
the highest q parameters (i.e., the brightest sources in the FIR
relative to the radio) would fall well above $q=3$.

One way of obtaining lower $q$-values, as has been pointed out several
times before \citep[e.g.,][]{vlahakis07,bourne11}
is through radio emission from a central AGN. This has an
interesting theoretical aspect, since it has been proposed
\citep[e.g.,][]{silk09} that
gas compression caused by expanding radio sources in massive
high-$z$ galaxies may in fact contribute to boosting the
star-formation efficiency in intense high-$z$ starbursts.  However,
although contamination with nuclear radio emission may be a particular
concern for stacking analyses (where one or a few relatively bright
radio sources may affect the average result of the overall sample), in
our case this explanation seems to be less likely. First, the radio sources
are spatially resolved in FIRST, suggesting they are extended, just
like the star formation. Even if the Dusty GEMS have extended radio
sources, the gravitational lensing would need to amplify the radio
emission by similar amounts to that of the star formation in most of our
sources; this appears relatively unlikely, given that the AGN radio
morphologies of high-$z$ galaxies are very different from those of the
dust and stellar components \citep[e.g.][]{sajina07}.

As a second counter argument,
our SED fits have already disfavoured the presence of very
luminous AGN. Although galaxies are known to host central radio
sources without bright bolometric emission, the host galaxies of such
AGN tend to have little dust, gas, and on-going star
formation. And although relatively faint AGN may be present without
leaving strong signatures in the dust SEDs, only a subset of those
should be radio loud. Even optimistic results \citep[e.g.,
  from][]{sajina07}, imply that only up to about 30\,\% of dusty
high-$z$ starburst galaxies may host moderately bright ($P_{\rm
  1.4}\approx10^{25}\,{\rm W}\,{\rm Hz}^{-1}$, comparable to our
sources for magnification factors of order 10) nuclear radio sources
that are as bright or brighter than the star formation itself. In
addition, these sources are classified as AGN in the mid-infrared. We
note that we have not found any obvious trends between the
$q$-parameter and possible probes of AGN contamination like dust
temperature or $L_{\rm FIR}$.  We also do not find a correlation
between the flux density at 22\,$\mu$m and the 1.4~GHz radio emission.

As an alternative, the offsets towards larger and smaller $q$-parameters
may be caused by the star-forming environments
themselves. \citet{lacki10} suggest that enhanced synchrotron emission
from cosmic rays in star-forming galaxies at high redshift could be
one outcome of the strong turbulence observed in these galaxies
\citep[e.g.,][]{fs09,lehnert09,swinbank11}, which enhances the scale
height of the gas, and lowers their volume density. In this case,
energy loss of cosmic rays through synchrotron radiation,
bremsstrahlung, and other processes could be either enhanced or
diminished, depending on the local magnetic fields and density
distribution of the ISM; this could either decrease or increase the
energy losses of the cosmic rays, making our sources either brighter
or fainter \citep[e.g.,][]{murphy09}.
Other possible explanations, which might also involve the
sources with unusually high $q$-parameters, include evolution in the
dust properties, the age of the starburst \citep[the radio should not
probe starbursts with ages of less than a few times $10^7$\,years or
greater than a few times $10^8$\,years, because of the timing of
core-collapse supernovae,][]{bressan02}, and a top-heavy initial
mass function \citep[e.g.,][]{baugh05}. It will be interesting to
obtain high-resolution centimetre-wave maps of these sources, along
with our gas and dust interferometry, to  constrain the potential
 impact of differential lensing and to  further elucidate how the
$q$-parameter depends locally on the ISM properties in our sources.

\section{Gas masses and gas-to-dust ratios}
\label{sec:dustgasratios}

To determine the molecular gas mass (which is dominated by H$_2$) from
the CO luminosity, we have to assume an empirical conversion factor,
which is notoriously difficult to justify from first principles, and
which is therefore still heavily debated in the literature, in
particular for high-redshift galaxies \citep[e.g.,][]{daddi10,
 genzel10, glover11, narayanan12}. The canonical value adopted for
most high-$z$ galaxies is $\alpha
\approx 0.8\,{\rm M}_{\odot}/({\rm K}\,{\rm km}\,{\rm s}^{-1}\,{\rm
 pc}^2$). This value was first derived by \citet{downes98} for the
dense, circumnuclear, optically thick molecular gas discs in nearby
ULIRGs, and is commonly adopted also for dusty starburst galaxies in
the early Universe. 

However, several studies in recent years (starting with
\citealt{daddi10} and \citealt{genzel10}) have called into question
whether a single CO(1-0)-to-H$_2$ conversion factor may be appropriate
to use for all high-z galaxies. Several attempts have therefore been
undertaken to constrain $\alpha_{\rm CO}$ either from dynamical mass
estimates, or on theoretical grounds. One empirical approach suited
for galaxies with well constrained dust mass measurements like ours,
is to assume that high-$z$ galaxies fall onto a similar relationship
between gas-to-dust mass ratios and metallicities as found for SINGs
galaxies in the local Universe \citep[][]{leroy08}. \citet{magdis11}
used this approach to confirm their $\alpha_{\rm CO}$ determinations,
which they previously obtained from dynamical mass estimates.

We can use our measurements of the dust mass, $M_d$ (Sect.~\ref{ssec:dust}
and Table~\ref{tab:dust}), and the CO line luminosity, $L^{\prime}$
(Sect.~\ref{ssec:lineprofiles} and Table~\ref{tab:lines}), to estimate
ratios of $L^{\prime}/M_d$, which scale with gas-phase metallicity
\citep[][]{leroy08,magdis11}. Using Fig.~3 of \citet{magdis11}, we
find that our measured range $L^{\prime}/M_d =40-140$ corresponds to
gas-phase metallicities $12+log(O/H)\sim 8.9-9.3$. These values are
appropriate for $\beta=2.0$ (Sect.~\ref{ssec:dust}).  Furthermore, we
adopted a ratio $r_{32/10}=1$ to convert from the luminosities of the
observed mid-J CO lines to CO(1-0). This factor is expected for
optically thick gas \citep[e.g.,][]{solomon05}. In the Cosmic Eyelash,
\citet{danielson11} measured $r_{32/10}=0.7$.

High gas-phase metallicities correspond to small values of
$\alpha_{\rm CO}$ of $\le 1.0$. For example, if we use the linear fit
between $\alpha_{CO}$ and metallicity of \citet{genzel12}, we find
conversion factors of about 0.4 at face value, although with large
uncertainties. Likewise, Fig.~3 of \citet{magdis11} suggests
$\alpha_{\rm CO}<1.0\,{\rm M}_{\odot}/({\rm K}\,{\rm km}\,{\rm
  s}^{-1}\,{\rm pc}^2$) for galaxies with $L^\prime/M_d$ ratios and
metallicities as high as in our sources. This suggests that using the
ULIRG conversion factor of $\alpha_{\rm CO} =0.8 \,{\rm
  M}_{\odot}/({\rm K}\,{\rm km}\,{\rm s}^{-1}\,{\rm pc}^2$) is more
appropriate than much higher factors of $3-5 \,{\rm
  M}_{\odot}/({\rm K}\,{\rm km}\,{\rm s}^{-1}\,{\rm pc}^2$), as
previously adopted for more moderately star forming, disk-like
high-redshift galaxies and the Milky Way. We stress that these results
are measured in small regions of high-z galaxies, and are not
necessarily representative of the average values in these galaxies
\citep[][]{serjeant12}.

When using a common conversion factor $\alpha_{\rm CO}\approx0.8\,{\rm
  M}_{\odot} /({\rm K}\,{\rm km}\,{\rm s}^{-1}{\rm pc}^2)$, we find
molecular gas masses of 2--10$\times 10^{11}\mu^{-1}{\rm M}_{\odot}$
for \Planck's Dusty GEMS (including the gravitational magnification
factor $\mu$). Results for individual galaxies are listed in 
Table~\ref{tab:lines}.

\section{Integrated star-formation law}
\label{sec:SK}

The tight correlation between molecular gas mass surface density and
star-formation intensity over scales of around 100\,pc to entire
galaxies highlights how star formation depends on the available
molecular gas reservoirs \citep[i.e., the Schmidt-Kennicutt
  law,][]{schmidt59,kennicutt98}, although there is no consensus about
a unique physical mechanism putting this relationship in place.

However, starting with, e.g., \citet{daddi10} and \citet{genzel10},
several studies have emphasized in recent years that not all
high-redshift galaxies may strictly obey the same empirical
star-formation law, but that at a given gas surface density, starburst
galaxies may be more efficient in turning their gas into stars. The
reasons for this are still unclear, with possibilities ranging from
changes in the stellar initial mass function \citep[][]{baugh05} to
different star-formation efficiencies. For example, there may be more
star-forming clouds in the ISM than in more quiescent high-$z$
galaxies \citep[][]{lehnert13}, or the star-formation efficiency per
free-fall time could be higher \citep[][]{dekel09}. What we can
confidently assert (given the absence of dust with $T\gg50\,$K, faint
22-$\mu$m flux density, or excess radio emission) is that AGN do not
play a dominant role in boosting the FIR continuum in the strongly
amplified regions that we are seeing in our galaxies
(Sect.~\ref{sec:templatefit}). Given the close astrophysical
connection between dust and gas in star-forming regions, we see no
reason to believe that differential lensing plays an important role
for our analysis of the gas conditions in the regions, which are
amplified by the gravitational lenses. Of course, we must keep in mind
that the gas and dust measurements obtained in small, selectively
amplified regions of these galaxies are not necessarily representative
of the average gas and dust properties in these galaxies on global
scales. \citep[][]{serjeant12}.

The relationship of far-infrared luminosity with CO line luminosity is
a simple, empirical, and robust way of investigating this
``star-formation law.''  It is directly related to the integral form of
the Schmidt-Kennicutt law, because the far-infrared
luminosity probing the emission from warm dust is an excellent tracer of star
formation. CO line emission is mainly excited through collisions with
H$_2$ and is the most commonly adopted proxy of the total molecular gas mass
\citep[e.g.,][]{omont07}.

\begin{figure}
\centering
\includegraphics[width=0.50\textwidth]{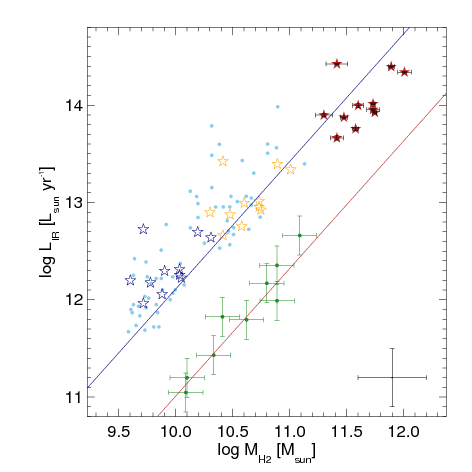}
\caption{{\it Left}: Infrared luminosities ($L_{\rm IR}$) as a
function of molecular gas mass for our sources.  Red, yellow, and
purple stars indicate the measured values (uncorrected for
gravitational amplification), and the same values corrected by
fiducial lensing factors of 10 and 50, respectively. In agreement with the
 dust-to-gas ratios discussed in Sect.~\ref{sec:dustgasratios} we adopt the
 classical ``ULIRG'' conversion factor $\alpha_{\rm CO}=0.8\,{\rm
 M}_{\odot} /({\rm K}\,{\rm km}\,{\rm s}^{-1}\,{\rm pc}^2)$ for
all sources. The
blue and red line indicate the relationships for starbursts and
``main-sequence'' galaxies emphasized by \citet{daddi10}.  Small
dots show the low- and high-$z$ samples of \citet{daddi10} for
comparison. Light blue dots along the upper line show low- and
high-$z$ ULIRGs and SMGs, respectively. The filled dark green
dots (with error bars) are $BzK$ galaxies at $z\sim2$; for details
of these samples see \citet{daddi10}. All error bars include the
measurement uncertainties. The black cross in the lower right corner
of the plot shows a fiducial factor 2 in statistical
uncertainties.}
\label{fig:lgasldust}
\end{figure}

In order to investigate whether our sources are representative
of star formation in high-$z$ starburst galaxies or more ordinary, but
nonetheless very intensely star-forming galaxies on the ``main
sequence'' \citep[e.g.,][]{elbaz11}, we use the molecular gas masses
estimated from the CO luminosities ($M_{\rm H2}$) and far-infrared
luminosities ($L_{\rm FIR}$), shown in Fig.~\ref{fig:lgasldust}, and
compare with the correlations found by \citet{daddi10} and
\citet{genzel10} for starburst galaxies, together with the more gradually, but
nonetheless intensely star-forming high-$z$ galaxies on the ``main
sequence.''  For this comparison, we use the integrated star-formation
rates and molecular gas masses, adopting
$\alpha_{\rm CO}=0.8\,{\rm M}_{\odot}/
 ({\rm K}\,{\rm km}\,{\rm s}^{-1}{\rm pc}^2)$ as justified in Sect.~\ref{sec:dustgasratios}. 

We find that all sources fall closer to the the upper ``starburst''
than the lower ``main sequence'' lines in Fig.~\ref{fig:lgasldust}. To
illustrate the effects of lensing, i.e., how these sources might
appear in the image plane, we plot not only the observed luminosities
(red filled stars), but also the luminosities for fiducial
magnification factors ${\mu}=10$ and ${\mu}=50$ (orange and light blue
empty stars in Fig.~\ref{fig:lgasldust}).  Even without the detailed
lens modelling their position in the diagram relative to typical
starburst or main-sequence galaxies does not depend sensitively on the
precise magnification factor.

We stress that in the
present study, we only show integrated measurements, whereas the
relationship that underlies the Schmidt-Kennicutt diagram is
between the surface density of molecular gas mass and the
star-formation rate. Our on-going interferometric follow-up programme
will enable us to derive more detailed constraints, and to investigate
whether the ``starburst'' and ``main sequence'' classifications are unique
for each source, or whether each individual Dusty GEM will show a range
in star-formation efficiency.

FIR luminosities and gas mass estimates also provide simple, rough
constraints on the gas depletion timescale, $t_{\rm dpl}=M_{\rm
  gas}$/SFR, and hence the time during which the current star
formation intensity can be maintained without replenishing the
molecular gas reservoirs. Assuming that star-formation rates are
constant, we find short gas depletion timescales, $t_{\rm
  dpl}=(0.5{-}6)\times 10^7$ years, significantly less than the
typical stellar age of a far-IR or submm galaxy \citep[a few times
  $10^8$ years,][]{smail04, lapi12}, and also somewhat, but not
dramatically, shorter than the gas depletion timescales found for
unlensed submm galaxies \citep[e.g.,][]{greve05}.  This highlights
again the fact that our sources have all the hallmarks of being
``ordinary'' dusty starburst galaxies placed under particularly
powerful cosmic microscopes.

\section{Summary}
\label{sec:summary}
We have presented a first analysis of an extensive multi-wavelength
follow-up campaign of a new sample of the brightest high-redshift
FIR/submm galaxies, discovered through the unique synergy of the
\Planck\ and \textit{Herschel\/} satellites.  \Planck's all-sky nature
and multi-frequency coverage allows us to select rare peaks in the
submm background and {\it Herschel\/} observations lead to sub-sample
of strongly lensed candidates -- \Planck's Dusty GEMS.  Their FIR
  peak flux densities are up to $S_{\rm 350}=1130$\,mJy at
  350$\,\mu$m, including six sources that are above the completeness
  limit of \Planck\ at the highest frequencies. Our sample extends
the very successful searches for gravitationally lensed high-$z$
galaxies already carried out with \textit{Herschel\/} and the SPT
towards the brightest, rarest targets on the FIR/submm sky, which
emphasized the need for a genuine all-sky survey to systematically
probe such exceedingly uncommon sources.

All sources in our sample are bright, isolated point sources in
SPIRE 250-$\mu$m maps (18\arcsec\ FWHM), and have the typical
FIR-to-mm SEDs of dusty, intensely star-forming galaxies at high
redshift. They have redshifts in the range $z=2.2$--3.6, based on
multiple bright millimetre emission lines obtained with EMIR at the
IRAM 30-m telescope. Their dust and gas properties provide firm
evidence that they are indeed gravitationally lensed galaxies, as is
further supported through interferometric observations of their dust
and gas morphologies, already obtained for most sources.

We used the \textit{WISE\/} survey at 22$\,\mu$m and the 1.4\,GHz VLA
FIRST survey to show that the far-infrared continuum of the
Dusty GEMS is not dominated by the radiation of powerful AGN. In
particular, we find that buried quasars cannot make a dominant
contribution to their observed FIR SEDs. All SEDs are well fitted with
a single modified blackbody distribution with temperatures $T_{\rm
  d}=33$--50\,K, covering the range of high-redshift starburst
galaxies, as well as more gradually, but still intensely star-forming,
high-$z$ galaxies on the ``main sequence.''  They show a wide
  scatter about the local far-infrared radio correlation, with
  $q$-parameters ranging from 2.0, as has previously been found for
  high-z galaxies by some authors, to above 2.7, which suggests a
  considerable excess of FIR relative to synchrotron emission. One
plausible interpretation is that this is probably a consequence of
their turbulent ISM, but this needs to be confirmed through more
detailed studies comparing the resolved radio emission with other
source properties. All galaxies have gas-to-dust ratios of
  $40-140$, consistent with a low CO-to-H$_2$ conversion factor, as
  expected for massive, dusty starburst galaxies with metallicities
  above solar. A full analysis of the spatially resolved properties
of these galaxies, as well as detailed lens modeling, is on-going.

Strongly lensed high-redshift submm galaxies represent an excellent
opportunity to study gas heating and acceleration, and the mechanism
driving star formation in the most vigorous starbursts in the early
Universe. Detailed observations of the dust, stellar populations, and
multiple emission and absorption lines, in particular with submm and
millimetre interferometry have already been obtained and will be
discussed in future papers.

\section*{Acknowledgements}
We would like to thank the staff at the IRAM 30-m telescope, in
particular N.~Billot and S.~Trevino for their excellent support during
observations. We are also very grateful to the former director of
IRAM, P.~Cox, the director of the SMA, R.~Blundell, the director of
the CFHT, D.~Simons, and the director of ESO, T.~de~Zeew, for the
generous allocation of Director's Discretionary Time. We thank the
referee, S. Bussman, for constructive comments that helped improve our
manuscript. We would also like to thank A.~Sajina and several other
colleagues unknown to us and sollicited by the Planck collaboration as
external referees for their valuable comments on an earlier version of
the paper. We would also like to thank C. Kramer for having made his
{\tt CLASS} routine FTSPlatformingCorrection5 available to us. We
thank the Planck Editorial Board for ensuring that our manuscript is
in accordance with the internal Planck publication rules and
standards.  MN acknowledges financial support from ASI/INAF agreement
2014-024-R.0 and from PIRN-INAF 2012 project ``Looking into the
dust-obscured phase of galaxy formation through cosmic zoom lenses in
the \textit{Herschel\/} Astrophysical Large Area Survey.'' IFC, LM and
EP acknowledge the support of grant ANR-11-BS56-015.

The largest part of this work is based on observations carried out with the
IRAM 30-m Telescope and the IRAM Plateau de Bure Interferometer.  IRAM
is supported by INSU/CNRS (France), MPG (Germany) and IGN (Spain).

The Submillimeter Array is a joint project between the Smithsonian
Astrophysical Observatory and the Academia Sinica Institute of Astronomy and
Astrophysics and is funded by the Smithsonian Institution and the Academia
Sinica.

Based in part on observations obtained with MegaPrime/MegaCam, a joint
project of CFHT and CEA/DAPNIA, at the Canada-France-Hawaii Telescope
(CFHT), which is operated by the National Research Council (NRC) of
Canada, the Institute National des Sciences de l'Univers of the Centre
National de la Recherche Scientifique of France, and the University of
Hawaii. Based in part on observations obtained with WIRCam, a joint
project of CFHT, Taiwan, Korea, Canada, France, and the
Canada-France-Hawaii Telescope (CFHT).

The development of \Planck\ has been supported by: ESA; CNES and
CNRS/INSU-IN2P3-INP (France); ASI, CNR, and INAF (Italy); NASA and
DoE (USA); STFC and UKSA (UK); CSIC, MICINN, JA, and RES (Spain);
Tekes, AoF, and CSC (Finland); DLR and MPG (Germany); CSA (Canada);
DTU Space (Denmark); SER/SSO (Switzerland); RCN (Norway); SFI
(Ireland); FCT/MCTES (Portugal); and PRACE (EU). A description of
the Planck Collaboration and a list of its members, including the
technical or scientific activities in which they have been involved,
can be found at
\url{http://www.rssd.esa.int/index.php?project=PLANCK&page=Planck_Collaboration}

The \textit{Herschel\/} spacecraft was designed, built, tested, and launched
under a contract to ESA managed by the \textit{Herschel}/\Planck\ Project team
by an industrial consortium under the overall responsibility of the prime
contractor Thales Alenia Space (Cannes), and including Astrium
(Friedrichshafen) responsible for the payload module and for system
testing at spacecraft level, Thales Alenia Space (Turin) responsible
for the service module, and Astrium (Toulouse) responsible for the
telescope, with in excess of a hundred subcontractors.


We acknowledge the use of the Galaxies, INterstellar mater \&
COsmology (GINCO) archive for the Integrated Data \& Operation Centre
(IDOC) at Institut d'Astrophysique Spatiale and Observatoire des
Sciences de l'Univers de l'Universit\'e Paris Sud (OSUPS). Support for
IDOC is provided by CNRS \& CNES. 

\bibliography{lens}

\end{document}